\documentclass[11pt]{article}

\usepackage[a4paper,margin=0.9in]{geometry}
\usepackage[authoryear,round]{natbib}
\usepackage{amsmath,amssymb,amsbsy,amsfonts,amsthm}
\usepackage[T1]{fontenc}
\usepackage[utf8]{inputenc}
\usepackage{microtype}
\usepackage{graphicx}
\usepackage{booktabs,multirow,array,calc}
\usepackage{caption}
\usepackage{subcaption}
\usepackage[hidelinks]{hyperref}
\usepackage{xurl}

\hypersetup{
  pdftitle={Homogeneity Pursuit for 3D Point Clouds},
  pdfauthor={Haojie Dong; Hanbing Zhu; Xinghui Wang; Xuejun Wang; Xinyuan Song},
  pdfkeywords={cluster analysis; latent groups; 3D point clouds; triangulation; trivariate splines}
}

\newtheorem{theorem}{Theorem}

\newtheorem{corollary}[theorem]{Corollary}

\theoremstyle{definition}

\theoremstyle{remark}
\newtheorem{remark}{Remark}

\title{Homogeneity Pursuit for 3D Point Clouds}
\author{
Haojie Dong\textsuperscript{1},
Hanbing Zhu\textsuperscript{1}\thanks{Corresponding author: \texttt{zhuhbecnu@163.com}.},
Xinghui Wang\textsuperscript{1}\thanks{Corresponding author: \texttt{wangxinghui@ahu.edu.cn}.},
Xuejun Wang\textsuperscript{1}, and
Xinyuan Song\textsuperscript{2}\\[1ex]
\small \textsuperscript{1}School of Big Data and Statistics, Anhui University,
Hefei, Anhui, China\\
\small \textsuperscript{2}Department of Statistics and Data Science,
The Chinese University of Hong Kong, Hong Kong, China
}
\date{}

\begin{document}
\maketitle

\begin{abstract}
This paper proposes a novel classification learning approach to identify the latent group structures in three-dimensional (3D) point clouds collected with irregular shapes, which allows 3D point clouds to be classified into different groups with the number of groups and the group membership unknown. 3D point clouds within a group have a common mean function, while mean function heterogeneity is allowed across groups. We first obtain preliminary estimates of the individual mean functions by penalized trivariate splines over triangulation. Based on these preliminary estimates, we develop an agglomerative clustering algorithm to estimate the unknown group structures and propose a ratio criterion to determine the number of groups. We establish the consistency of the estimated number of groups and group membership. Furthermore, we investigate a post-grouping trivariate penalized spline estimation procedure for group-specific mean functions and derive the corresponding asymptotic theory. The proposed method is validated through simulation studies and showcased with an application to the imaging data from the Alzheimer's Disease Neuroimaging Initiative database.
\end{abstract}

\vspace{0.5em}

\section{Introduction}\label{sec1}

\noindent With recent advances in data acquisition technology, 3D point cloud data have become increasingly common in fields such as computer vision, medical imaging, and autonomous driving (\cite{lee20163d}, \cite{ruan2020review} and \cite{wijaya2024enhancing}). 3D point cloud data are usually collections of a vast number of measurements in space defined by a given coordinate system. They provide rich insights into the structure, geometry, and characteristics of real-world objects and environments. Efficiently extracting meaningful information from 3D point cloud data has become increasingly important across multiple applications. To support the visualization and interaction of large-scale point clouds, various methods such as multi-resolution point rendering systems and hierarchical point cloud structures have been proposed (\cite{Levoy2000qsplat} and \cite{gobbetti2004layered}). Methods for reconstructing continuous geometric surfaces from discrete point clouds have been systematically developed (\cite{berger2017survey}). In addition, nonparametric regression is widely used to model 3D point cloud data; see \cite{chung2018discrete}, \cite{li2024nonparametric}, \cite{wang2024efficient} and \cite{wang2025statistical}.

A typical assumption in existing nonparametric regression for 3D point cloud data is that the regression functions are homogeneous across all 3D point clouds. However, heterogeneity is intrinsic to 3D point cloud data, as they typically originate from diverse objects and acquisition conditions. For instance, \cite{lin2024latent} identified latent groups in 3D functional magnetic resonance imaging (MRI) data from the Adolescent Brain Cognitive Development study. In this paper, we study nonparametric regression with latent group structures for 3D point clouds. An important advantage of latent group structure models is that, unlike completely heterogeneous or fully homogeneous models, they allow 3D point clouds to be classified into different groups, where individuals within a group share the same regression function, while heterogeneity exists across groups. Since the group membership and number of groups are unknown in these models, determining the true number of groups and group membership is a key question.

\begin{figure}
\centering
\begin{minipage}{0.23\textwidth}\centering\includegraphics[width=\linewidth]{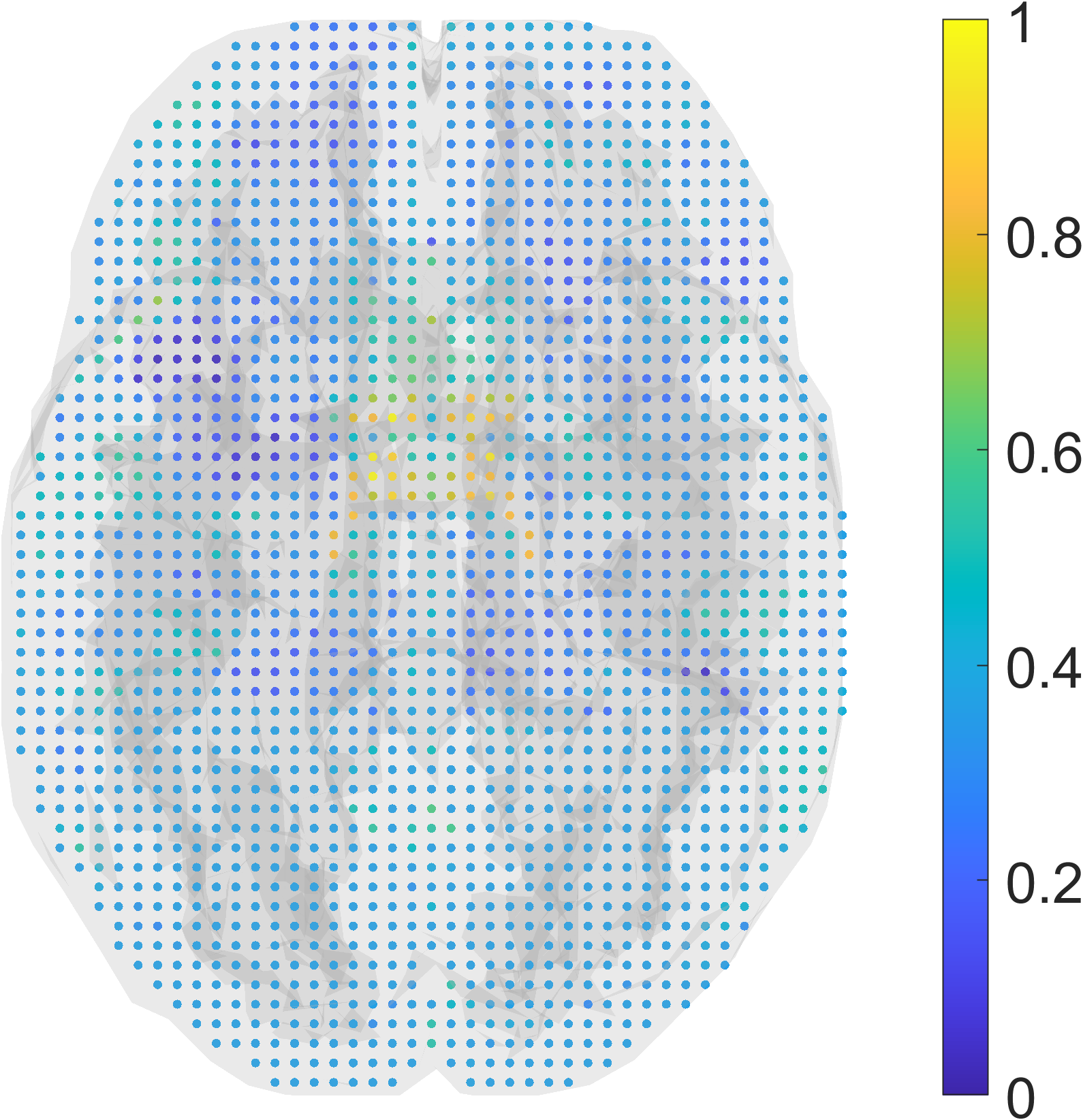}\captionsetup[subfigure]{labelformat=empty}\subcaption{}\end{minipage}
\hspace{0.05\textwidth}
\begin{minipage}{0.23\textwidth}\centering\includegraphics[width=\linewidth]{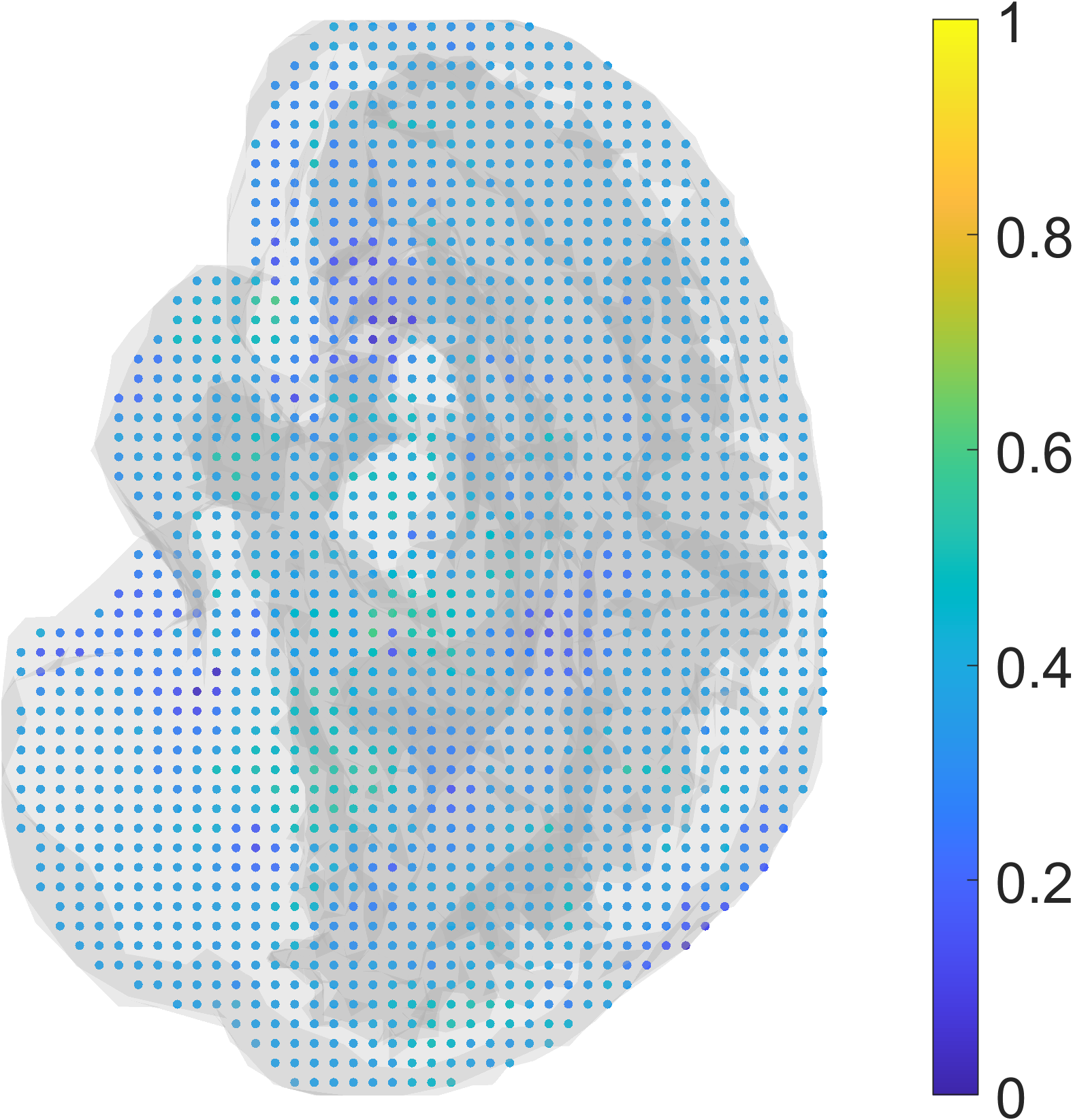}\captionsetup[subfigure]{labelformat=empty}\subcaption{}\end{minipage}
\hspace{0.05\textwidth}
\begin{minipage}{0.23\textwidth}\centering\includegraphics[width=\linewidth]{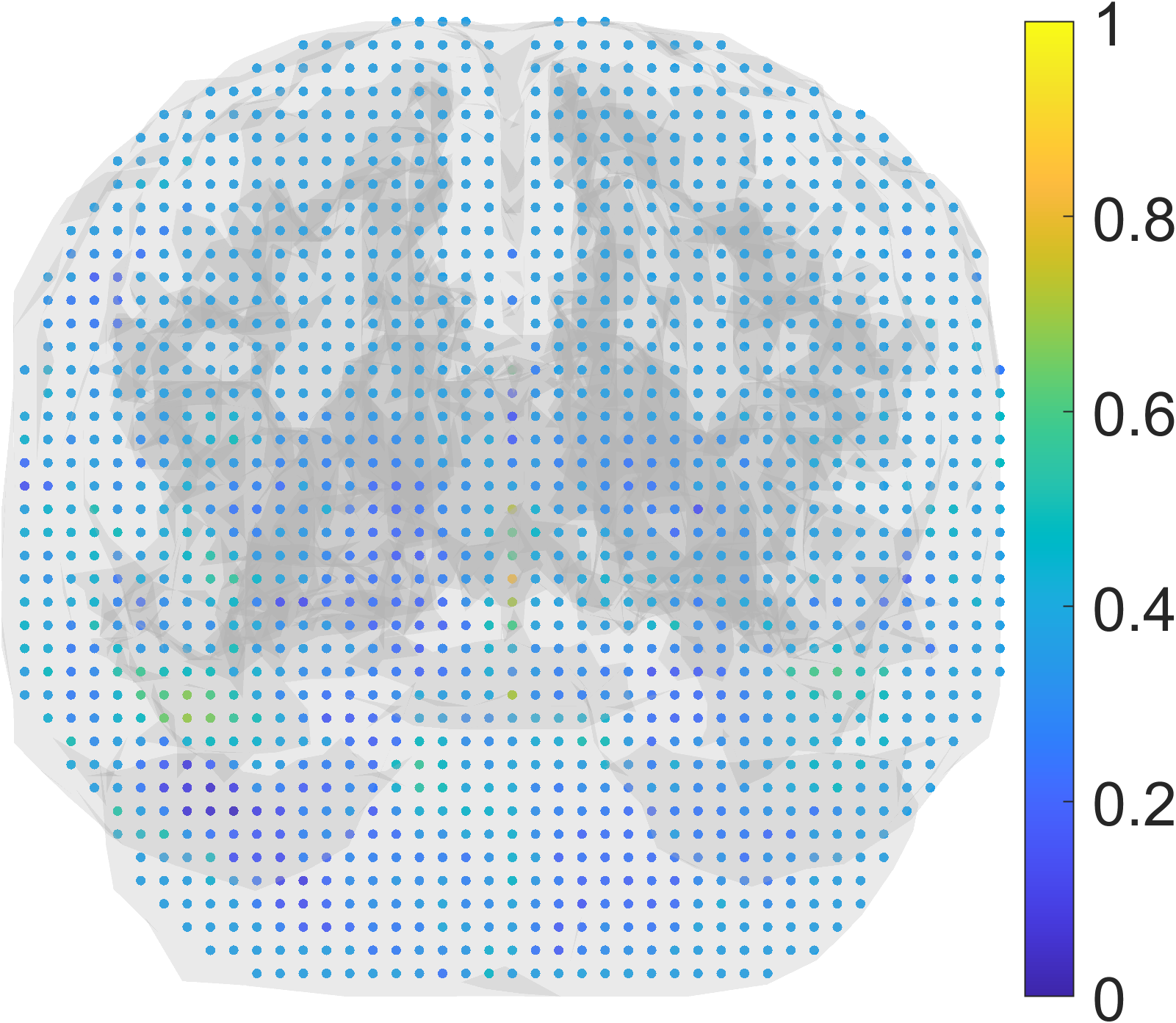}\captionsetup[subfigure]{labelformat=empty}\subcaption{}\end{minipage}
\caption{Observed brain point cloud for one ADNI subject from different viewing angles. The gray translucent background is the standard reconstructed brain structure used as a spatial reference, while the colored points represent the observed normalized MRI intensity values.}\label{realbrain}
\end{figure}

Identification of latent group structures has been extensively studied. For example, \cite{ke2016structure,pei2018nonparametric,mehrabani2023estimation} investigated group identification in fixed effects panel data models; \cite{su2019sieve,pei2022network,wang2024panel} extended this framework to varying coefficient models; and \cite{leng2023multi} further generalized it to quantile regression models. \cite{vogt2017classification,vogt2020multiscale} explored classification problems related to nonparametric functions in longitudinal data models. \cite{guo2022homogeneity} examined group identification in semiparametric factor models, while \cite{xiao2021homogeneity} tested for the existence of group structures under heavy-tailed errors. Despite extensive research on identifying latent group structures, effective statistical methods for identifying latent group structures in 3D point clouds remain unexplored. The core challenge in statistical analysis for 3D point clouds lies in handling irregular domains. A representative example is the Alzheimer's Disease Neuroimaging Initiative (ADNI) 3D brain MRI point cloud data shown in Figure \ref{realbrain}, in which the observed intensity values are defined over an irregular brain region rather than on a regular grid.

Traditional smoothing methods, such as kernel smoothing (\cite{jo2007spatial}), tensor product smoothing (\cite{reiss2010functional}), and wavelet smoothing (\cite{morris2006wavelet}), often perform poorly when handling such irregular domains due to oversmoothing of boundary features while suffering from the ``leakage'' problem, which refers to inaccurate signal reconstruction on complex domains; see \cite{ramsay2002spline} and \cite{wang2007low}. To address the spatial smoothing problem over complex domains, triangulation emerges as an effective strategy, particularly suited for irregular domains with complex contours or internal holes. This method enables more precise capture of key morphological features in imaging data. Leveraging its flexibility, tunable smoothness, and good adaptability to irregular domains, this method has been widely applied in both 2D (\cite{wang2020simultaneous}, \cite{yu2021multivariate}, and \cite{li2021sparse}) and 3D (\cite{li2024nonparametric} and \cite{wang2024efficient}) biomedical image analysis.

This paper investigates the estimation and identification of latent group structures in 3D point cloud data over irregular domains, which remains an open challenge. We first propose a trivariate penalized spline estimation method to recover underlying signals from error-contaminated observations. As there is no prior information on the latent groups, we obtain a preliminary estimation of the subject-specific mean functions, using only the estimated underlying signals from one individual. According to the preliminary estimates of the mean functions, we compute the distance matrix between individuals and subsequently use an agglomerative clustering algorithm to estimate the unknown group structure for the heterogeneous mean functions. Then, we introduce a ratio criterion to estimate the group number. Furthermore, we propose a post-grouping trivariate penalized spline smoothing method to estimate the group-specific mean functions.

The main contributions of this paper are threefold. First, we propose a novel classification learning method to identify unknown group structures for 3D point clouds over irregular domains. We develop an efficient algorithm and a simple ratio criterion to determine the number of latent groups and the group membership, and introduce a post-grouping estimation procedure to estimate group-specific mean functions. Second, we provide theoretical foundations for the proposed method. In particular, we establish the consistency of the estimated number of groups and the group membership. Additionally, we investigate the consistency and asymptotic normality of the post-grouping estimators. Such a theoretical investigation has not been conducted in the existing literature. Finally, we apply the proposed method to analyze the MRI data from the ADNI study and obtain new insights into Alzheimer's Disease (AD) diagnosis, which is essential for early intervention and personalized medicine.

The rest of this paper is organized as follows. Section \ref{sec2} introduces the model setup, the agglomerative clustering algorithm, and the ratio criterion for estimating the latent group structures, as well as the post-grouping spline smoothing estimation for the group-specific mean functions. The technical assumptions and main asymptotic properties are provided in Section \ref{sec3}. Section \ref{sec4} presents simulation studies, including implementation details, to assess the finite-sample performance of the proposed methodology. The usefulness of the proposed methodology is further demonstrated by a real-data example in Section \ref{sec5}. Section \ref{sec6} concludes the paper. All technical proofs of the main results are provided in Appendix A.

\section{Model and Estimation Methodology}\label{sec2}

The domain of 3D point clouds is usually made up of irregular geometric domains $\Omega \in \mathbb{R}^3$. In the subsequent analysis, we consider a collection of $n$ 3D point clouds, the $i$-th 3D point cloud is conceptualized as a set of observations $\left\{(Y_{ij},\boldsymbol{p}_{j}),j=1,\ldots,N\right\}$, where $Y_{ij}$ is the response variable observed at the location $\boldsymbol{p}_{j}\in\Omega$. We assume that the 3D point clouds exhibit a true group structure $\left\{\mathcal{G}_1, \ldots, \mathcal{G}_{K_0}\right\}$, a partition of $\{1,2, \ldots, n\}$. To extract the underlying signal and identify homogeneous relationships of the 3D point clouds, we consider the following nonparametric regression model with the latent group structure:
\begin{align}
Y_{ij}&=\mu_i\left(\boldsymbol{p}_{j}\right)+ \varepsilon_{i}\left(\boldsymbol{p}_{j}\right),\quad \boldsymbol{p}_j\in\Omega,\label{model}\\
\mu_i(\cdot)&=g_k(\cdot) \text { for } i \in \mathcal{G}_k \text { and } \mathcal{G}_{k_1} \cap \mathcal{G}_{k_2}=\emptyset \text { for } 0\leq k_1 \neq k_2\leq K_0,\label{group}
\end{align}
where $g_k(\cdot)$ is the unknown group-specific mean function, $\mu_i(\cdot)$ is the subject-specific mean function, and random error $\varepsilon_{i}\left(\boldsymbol{p}_{j}\right)$ with mean zero and variance $\sigma^2$. The $\emptyset$ denotes the empty set. 

\subsection{Trivariate Spline over Triangulations}

We approximate nonparametric functions using trivariate spline basis functions over a triangulation, which are piecewise polynomial trivariate functions defined on tetrahedra. The detailed definitions of these basis functions are provided in Appendix A. Figure \ref{triangulation} illustrates examples of triangulation for several irregular domains.

\begin{figure}
\centering
\begin{minipage}{0.18\textwidth}\centering\includegraphics[width=\linewidth]{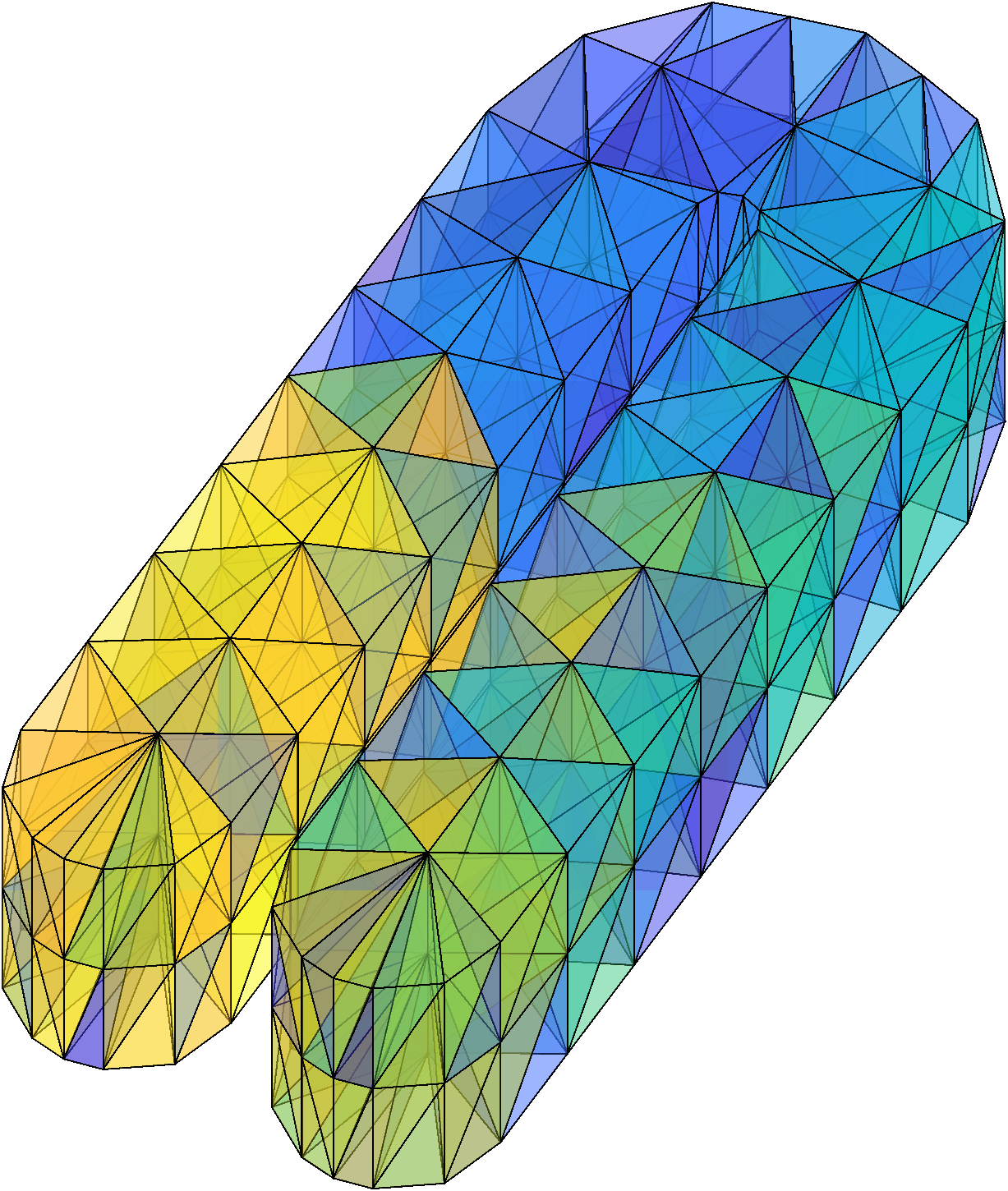}\captionsetup[subfigure]{labelformat=empty}\subcaption{}\end{minipage}
\hspace{0.05\textwidth}
\begin{minipage}{0.21\textwidth}\centering\includegraphics[width=\linewidth]{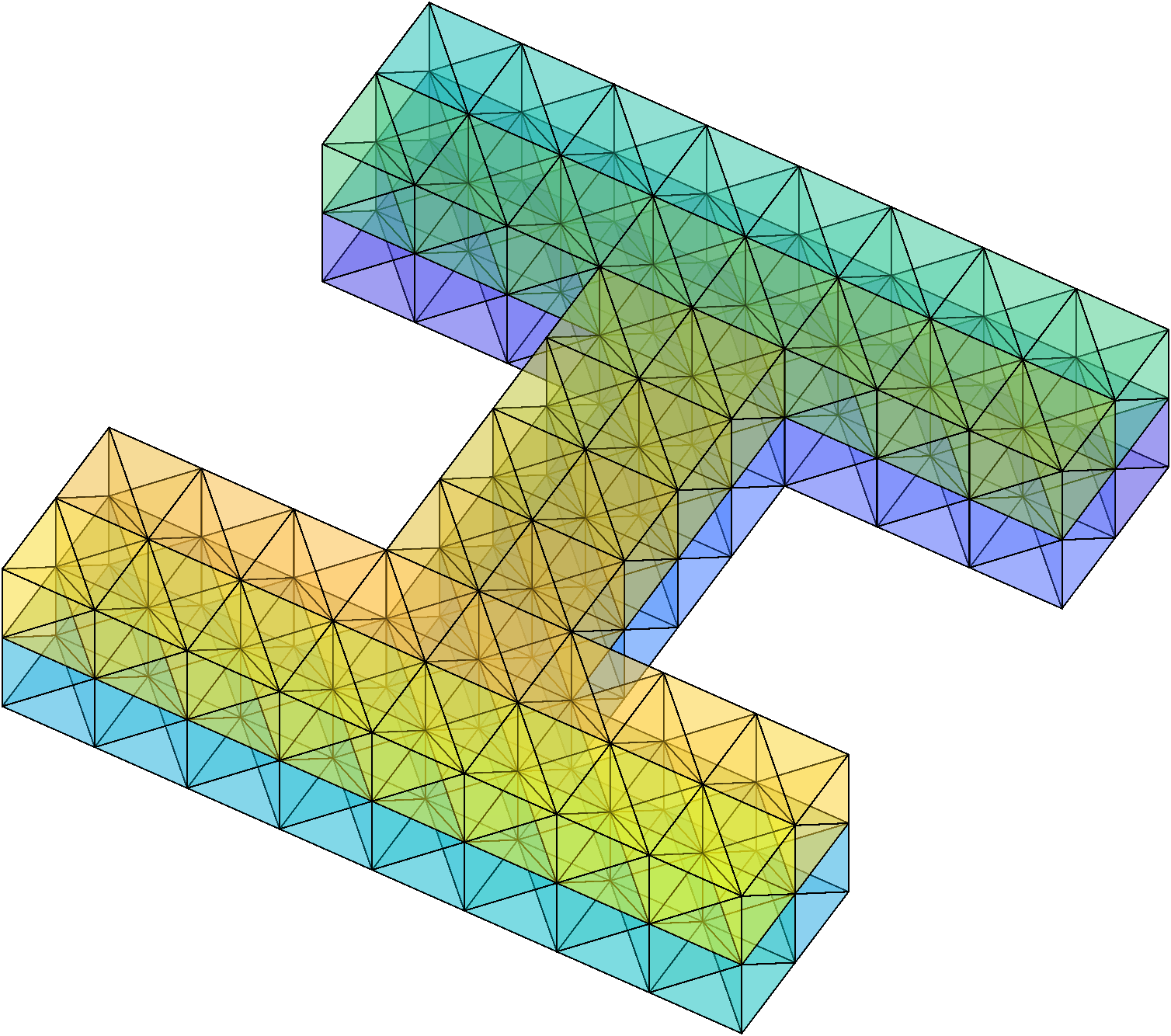}\captionsetup[subfigure]{labelformat=empty}\subcaption{}\end{minipage}
\hspace{0.05\textwidth}
\begin{minipage}{0.21\textwidth}\centering\includegraphics[width=\linewidth]{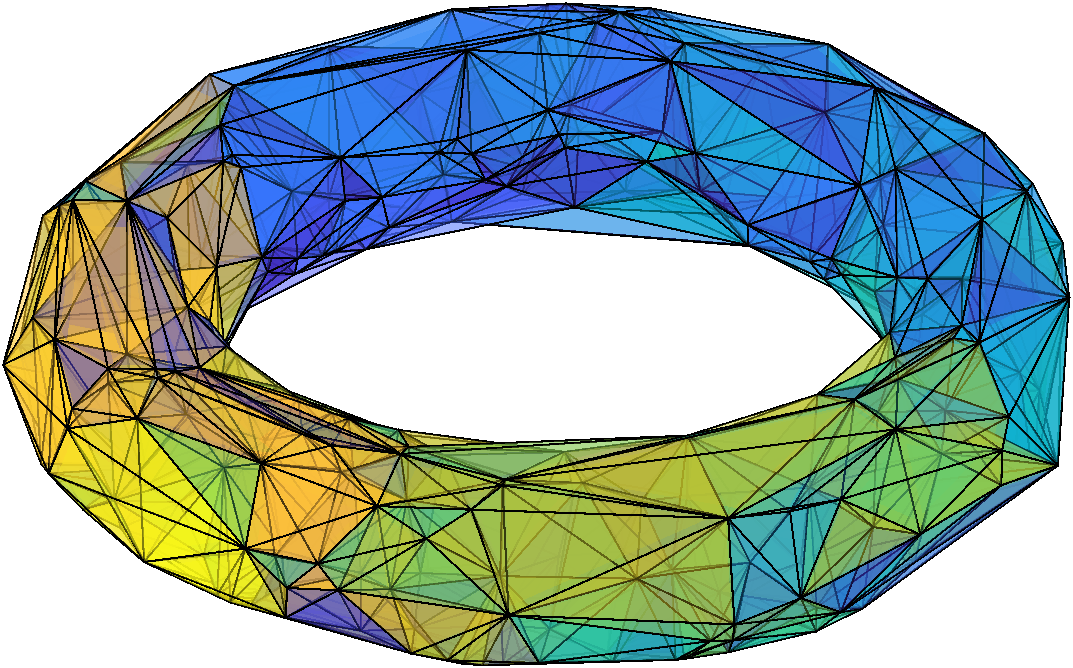}\captionsetup[subfigure]{labelformat=empty}\subcaption{}\end{minipage}
\hspace{0.05\textwidth}
\begin{minipage}{0.17\textwidth}\centering\includegraphics[width=\linewidth]{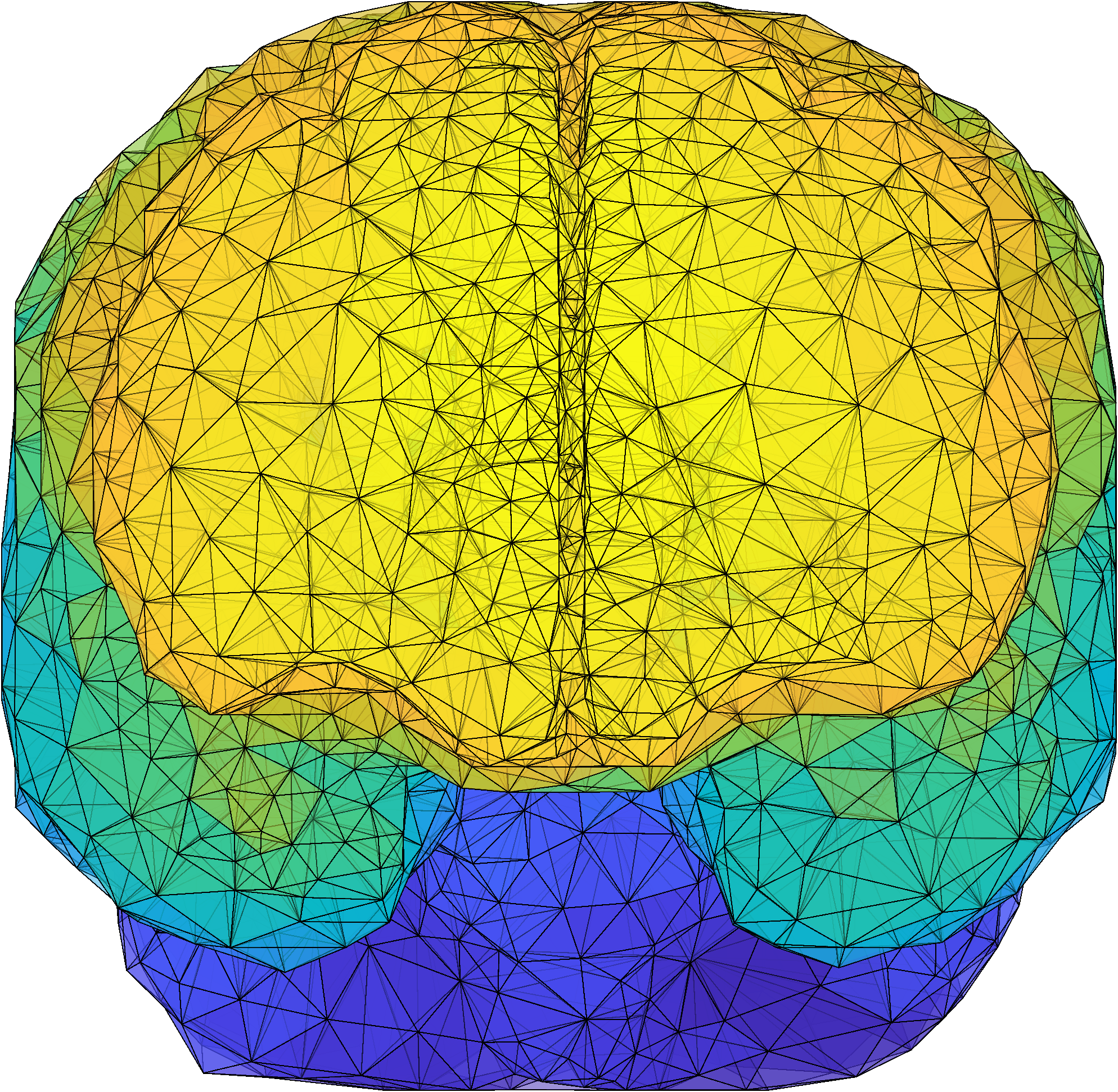}\captionsetup[subfigure]{labelformat=empty}\subcaption{}\end{minipage}
\caption{Example of triangulations of several irregular domains and the brain.}\label{triangulation}
\end{figure}

Consider a tight polyhedral domain, $\Omega$, which encompasses all the data points of a given 3D point cloud. The theoretical framework below treats $\Omega$ as a fixed, irregular, and predetermined entity. Denote a tetrahedron by $\tau$, the convex hull of four noncoplanar points in $\mathbb{R}^3$. Let $N_{\tau}$ be the number of tetrahedra within the domain $\Omega$. A collection $\triangle=\left\{\tau_1, \ldots, \tau_{N_{\tau}}\right\}$ of tetrahedra constitutes a triangulation of $\Omega=\cup_{h=1}^{N_{\tau}} \tau_h$, under the condition that any two tetrahedra in $\triangle$ intersect, if at all, only at a common vertex, along a common edge, or across a common triangular face.

Furthermore, let $|\tau|$ denote the length of the longest edge of the tetrahedron $\tau$ and $\varrho_\tau$ be the radius of the largest ball that can be inscribed within $\tau$. The ratio $\beta_\tau:=|\tau| / \varrho_\tau$ is called the shape parameter of $\tau$. The shape parameter $\beta_\tau$ characterizes the geometric shape of $\tau$, with a larger value of $\beta_\tau$ indicating a flatter tetrahedron. Utilizing this shape parameter, we can define that a triangulation $\triangle$ is \emph{$\beta$-quasi-uniform} if there is a positive value $\beta$ such that $\triangle$ satisfies: $|\triangle| / \varrho_\tau \leq \beta<\infty$, for any $\tau \in \triangle,$ where $|\triangle|:=\max \{|\tau|, \tau \in \triangle\}$ denotes the size of $\triangle$, which corresponds to the length of the longest edge within $\triangle$. 
%For a $\beta$-quasi-uniform triangulation, we have $N_{\tau} \leq\left(4 \pi|\triangle|^3\right)^{-1} 3 V_{\Omega} \beta^3$, where $V_{\Omega}$ denotes the volume of $\Omega$.

For any nonnegative integer $r$, we use $\mathcal{C}^r(\Omega)$ to denote the collection of all $r$-th continuously differentiable functions over $\Omega$. Let $\mathcal{S}_d^r(\triangle)=\left\{s \in \mathcal{C}^r(\Omega):\left.s\right|_\tau \in \mathcal{P}_d(\tau), \tau \in \triangle\right\}$ be a spline space of degree $d$ and smoothness $r$ over $\triangle$, where $\left.s\right|_\tau$ is the polynomial piece of spline $s$ restricted on tetrahedron $\tau$, and $\mathcal{P}_d(\tau)$ be the space of all trivariate polynomials defined on $\tau$ with total degrees less than or equal to $d$. Let $\left\{B_{\xi}(\cdot)\right\}_{\xi \in \mathcal{M}}$ be a set of Bernstein basis polynomials for $\mathcal{S}_d^r\left(\triangle\right)$, where $\mathcal{M}$ is the index set of all the spline basis functions. Then,  we can represent any spline function $s \in \mathcal{S}_d^r(\triangle)$ using the following basis expansion: 
\begin{equation}
s(\boldsymbol{p})=\sum_{\xi\in\mathcal{M}}B_{\xi}(\boldsymbol{p}) \gamma_{\xi} =\mathbf{B}^{\top}(\boldsymbol{p}) \boldsymbol{\gamma}, \label{sdr}
\end{equation}
where $\mathbf{B}(\boldsymbol{p})=\{B_{\xi}(\boldsymbol{p}),\xi \in \mathcal{M}\}^{\top}$ and spline coefficient vector $\boldsymbol{\gamma}=(\gamma_\xi, \xi \in \mathcal{M})^{\top}$.

To meet the smoothness requirement of the splines, we need to impose some constraints on the spline coefficients in \eqref{sdr}. The smoothness conditions are linear. We require that $\boldsymbol{\gamma}$ satisfies $\mathbf{H} \boldsymbol{\gamma}=0$ with $\mathbf{H}$ being the matrix for all smoothness conditions across the shared faces of the tetrahedron, which depends on the smoothness parameter $r$ and the triangulation structure; see \cite{awanou2006multivariate}, \cite{lai2007spline} and \cite{zhou2014smoothing}.

\subsection{Preliminary Penalized Spline Approximation over Triangulations}\label{sec2.1}
To obtain a preliminary estimate of the subject-specific mean function $\mu_i(\cdot)$, we consider the following trivariate penalized spline regression problem:
\begin{equation}
\min _{s_i \in \mathcal{S}_d^r(\triangle)} \sum_{j=1}^{N}\left\{Y_{ij}-s_i\left(\boldsymbol{p}_{j}\right)\right\}^2+h_i \mathcal{E}(s_i), \label{objective f}
\end{equation}
where $h_i>0$ is the roughness penalty parameter that controls the smoothness of the fitted spline function $s_i$. The energy functional is defined as 
\begin{equation}
\mathcal{E}(g)=\sum_{|\boldsymbol{\alpha}|=2}\binom{2}{\alpha_1}\binom{2-\alpha_1}{\alpha_2} \int_{\Omega}\left\{\nabla^{\boldsymbol{\alpha}} g(\boldsymbol{p})\right\}^2 d\boldsymbol{p}, \label{roughness penalty}
\end{equation}
in which for any function $f$ over the closure of domain $\Omega$ and any index $\boldsymbol{\alpha}=\left(\alpha_1, \alpha_2, \alpha_3\right)$ of order $|\boldsymbol{\alpha}|=\alpha_1+\alpha_2+\alpha_3$, the derivatives $\nabla^{\boldsymbol{\alpha}} f=\partial^{|\boldsymbol{\alpha}|} f / \partial_x^{\alpha_1} \partial_y^{\alpha_2} \partial_z^{\alpha_3}$.

According to \eqref{sdr}, \eqref{objective f} is equivalent to \begin{equation*}
\min _{\boldsymbol{\gamma}_i} \sum_{j=1}^{N}\left\{Y_{ij}-\mathbf{B}^{\top}\left(\boldsymbol{p}_{j}\right) \boldsymbol{\gamma}_i\right\}^2+h_i\boldsymbol{\gamma}_i^{\top} \mathbf{P} \boldsymbol{\gamma}_i,
\end{equation*} 
subject to $\mathbf{H} \boldsymbol{\gamma}_i=\mathbf{0}$, where $\mathbf{P}$ is the diagonally block penalty matrix satisfying that $\boldsymbol{\gamma}_i^{\top} \mathbf{P} \boldsymbol{\gamma}_i = \mathcal{E} (\mathbf{B}^\top \boldsymbol{\gamma}_i)$. To solve the constrained minimization problem, we employ QR decomposition to eliminate this constraint. Specifically, \begin{equation*}
\mathbf{H}^{\top}=\mathbf{Q}\mathbf{R}=\left(\mathbf{Q}_{1} \mathbf{Q}_{2}\right)\binom{\mathbf{R}_{1}}{\mathbf{R}_{2}},
\end{equation*}where $\mathbf{Q}$ is an orthogonal matrix, $\mathbf{R}$ is an upper triangular matrix, $\mathbf{R}_{1}$ is a full-rank matrix with the same rank as $\mathbf{H}$, and $\mathbf{R}_{2}$ is a zero matrix. Note that for any vector $\boldsymbol{\gamma}_i$ that satisfies $\mathbf{H} \boldsymbol{\gamma}_i=\mathbf{0}$, there exists some $\boldsymbol{\theta}_i$ such that $\boldsymbol{\gamma}_i=\mathbf{Q}_{2} \boldsymbol{\theta}_i$; see Lemma 1 in \cite{wang2020efficient}. Thus, the constrained minimization problem is reduced to the following unconstrained minimization problem:
\begin{equation}
\min_{\boldsymbol{\theta}_i}\sum_{j=1}^{N}\left\{Y_{ij}-\widetilde{\mathbf{B}}^{\top}\left(\boldsymbol{p}_{j}\right) \boldsymbol{\theta}_i\right\}^2+h_i\boldsymbol{\theta}_i^{\top} \mathbf{D} \boldsymbol{\theta}_i, \label{smooth object}
\end{equation}
where $\widetilde{\mathbf{B}}(\boldsymbol{p})=\mathbf{Q}_{2}^{\top} \mathbf{B}(\boldsymbol{p})$ and $\mathbf{D}=\mathbf{Q}_{2}^{\top} \mathbf{P} \mathbf{Q}_{2}$. Then the minimizer of \eqref{smooth object} is given by \begin{equation*}
\widehat{\boldsymbol{\theta}}_i=\left\{\sum_{j=1}^N\widetilde{\mathbf{B}}(\boldsymbol{p}_j)\widetilde{\mathbf{B}}^{\top}(\boldsymbol{p}_j)+h_i\mathbf{D}\right\}^{-1} \left\{\sum_{j=1}^N\widetilde{\mathbf{B}}(\boldsymbol{p}_j)Y_{ij}\right\}.
\end{equation*} Consequently, the subject mean estimators $\widehat{\mu}_i(\boldsymbol{p})=\widetilde{\mathbf{B}}^{\top}(\boldsymbol{p})\widehat{\boldsymbol{\theta}}_i$.

\subsection{Estimation of Group Membership}\label{sec2.2}
Let $\boldsymbol{\Pi}$ be an $n \times n$ distance matrix representing the distances between the true subject mean functions. The diagonal elements of $\boldsymbol{\Pi}$ are set to zero as they correspond to the distances between each mean function and itself. The off-diagonal elements $\Pi(i, i^{\prime}), 1 \leq i \neq i^{\prime} \leq n$, are defined as
\begin{equation*}
\Pi(i, i^{\prime})=\int_{\Omega}|\mu_i(\boldsymbol{p})-\mu_{i^{\prime}}(\boldsymbol{p})|d\boldsymbol{p}.
\end{equation*}
Given the preliminary estimates of the subject-specific mean functions, we can compute an estimate of $\Pi(i, i^{\prime})$ as follows:
\begin{equation*}
\widehat{\Pi}(i, i^{\prime})=\frac{1}{N} \sum_{j=1}^{N}|\widehat{\mu}_i\left(\boldsymbol{p}_{j}\right)-\widehat{\mu}_{i^{\prime}}\left(\boldsymbol{p}_{j}\right)|.
\end{equation*}
Subsequently, we can obtain $\widehat{\boldsymbol{\Pi}}$, an estimated $n \times n$ distance matrix of $\boldsymbol{\Pi}$. The $(i, i^{\prime})$-entry of $\widehat{\boldsymbol{\Pi}}$ is $\widehat{\Pi}(i, i^{\prime})$ and the diagonal elements of $\widehat{\boldsymbol{\Pi}}$ are zeros. With this estimated distance matrix, we can employ the agglomerative clustering method. Next, we will describe the clustering algorithm when the number of groups is assumed to be $K$.
\begin{itemize}
    \item STEP 1: Start with $n$ groups, each corresponding to one of the $n$ 3D point clouds. Search for the smallest off-diagonal element in $\widehat{\boldsymbol{\Pi}}$, which is the smallest distance estimate.
    \item STEP 2: Merge the two groups with the smallest distance. Consequently, the number of groups reduces from $n$ to $n-1$. Update the estimated distance matrix for the $n-1$ groups. Here, the distance between two groups $\mathcal{G}_1$ and $\mathcal{G}_2$ is calculated via the complete linkage, that is, compute the farthest distance between an element in $\mathcal{G}_1$ and that in $\mathcal{G}_2$.
    \item STEP 3: Repeat the previous steps with the updated distance matrix and stop the algorithm when the number of groups reaches $K$.
\end{itemize}

Consequently, given any $K$, the estimated groups $\widehat{\mathcal{G}}_{1}, \ldots, \widehat{\mathcal{G}}_{K}$  can be obtained by the clustering algorithm. Specifically, if the true number of groups, $K_0$, is known a priori, then denote the estimated groups $\widetilde{\mathcal{G}}_{1}, \ldots, \widetilde{\mathcal{G}}_{K_0}$. 

\subsection{Estimation of the Number of Groups}\label{sec2.3}

Next, we introduce a ratio criterion to estimate the number of groups consistently. For a given number $K$, using the estimated groups $\widehat{\mathcal{G}}_{1}, \ldots, \widehat{\mathcal{G}}_{K}$ obtained by the agglomerative clustering algorithm, we can pool the estimators $\widehat{\mu}_i(\cdot), i\in \widehat{\mathcal{G}}_{k}$, and calculate the following estimate:
\begin{equation*}
\widehat{g}_{k}(\boldsymbol{p})=\frac{1}{\left|\widehat{\mathcal{G}}_{k}\right|} \sum_{i \in \widehat{\mathcal{G}}_{k}} \widehat{\mu}_i(\boldsymbol{p}), \ k=1, \ldots, K,
\end{equation*}
where $|\mathcal{G}|$ denotes the cardinality of a set $\mathcal{G}$. Subsequently, we calculate the average deviation for $\widehat{\mu}_i(\cdot)$ if we assume the number of groups is $K$:
\begin{equation}
\mathrm{D}(K)=\frac{1}{N K} \sum_{k=1}^K \frac{1}{\left|\widehat{\mathcal{G}}_{k}\right|} \sum_{i \in \widehat{\mathcal{G}}_{k}} \sum_{j=1}^{N}|\widehat{\mu}_i\left(\boldsymbol{p}_{j}\right)-\widehat{g}_{k}\left(\boldsymbol{p}_{j}\right)|.
\label{criterion}
\end{equation}

\begin{remark} \label{DK}
Suppose that the true groups are $\left\{\mathcal{G}_1, \ldots, \mathcal{G}_{K_0}\right\}$. If we take $K=K_0$, the clustering algorithm will terminate when the iteration reaches the true number of groups $K_0$, and the model \eqref{group} fits well. According to Theorem \ref{group priori}, the estimated groups $\left\{\widetilde{\mathcal{G}}_1, \ldots, \widetilde{\mathcal{G}}_{K_0}\right\}$ are equal to $\left\{\mathcal{G}_1, \ldots, \mathcal{G}_{K_0}\right\}$ with probability approaching one. In this case, $\mathrm{D}(K_0)$ converges to zero. 

However, if taking $K>K_0$, the clustering algorithm will stop before reaching $K_0$. The model \eqref{group} is overfitted as some of $\left\{\mathcal{G}_1, \ldots, \mathcal{G}_{K_0}\right\}$ are further split into smaller groups. For example, $K=K_0+1$, the estimated groups are $\left\{\widehat{\mathcal{G}}_1, \ldots, \widehat{\mathcal{G}}_{K_0+1}\right\}$, which implies that one of $\left\{\mathcal{G}_1, \ldots, \mathcal{G}_{K_0}\right\}$ falsely splits into two groups. Still, in this case, $\mathrm{D}(K_0+1)$ also converges to $0$ (illustrated in the proof of Theorem \ref{group estimate}). 

In addition, if taking $K<K_0$, the clustering algorithm will continue to run after reaching $K_0$, and the model \eqref{group} is underfitted, as some of $\left\{\mathcal{G}_1, \ldots, \mathcal{G}_{K_0}\right\}$ are falsely merged. For example, $K=K_0-1$, the estimated groups are $\left\{\widehat{\mathcal{G}}_1, \ldots, \widehat{\mathcal{G}}_{K_0-1}\right\}$, which implies that two groups of $\left\{\mathcal{G}_1, \ldots, \mathcal{G}_{K_0}\right\}$ are incorrectly merged. In this case, $\mathrm{D}(K_0-1)$ will be strictly larger than a positive constant by Assumption (C7)(ii).
\end{remark}

It is important to note that whenever $K$ exceeds or equals $K_0$, the function $\mathrm{D}(K)$ consistently converges to zero. Consequently, to accurately determine $\widehat{K}$, additional technical approaches are required. We set $\mathrm{D}(1)/\mathrm{D}(0)=1$ and define $0/0\equiv 1$, then we can utilize the following ratio criterion to ascertain $\widehat{K}$:
\begin{equation}
\widehat{K}=\underset{1 \leq K \leq \overline{K}}{\arg \min } \frac{\mathrm{D}(K)}{\mathrm{D}(K-1)},
\label{ratio}
\end{equation}
where $\overline{K}$ is a pre-specified positive integer larger than $K_0$, and we set $\mathrm{D}(K)=0$ if $\mathrm{D}(K)$ is smaller than $\varpi$, a theoretical threshold satisfying some mild restrictions. 

\subsection{Post-grouping Penalized Spline Approximation over Triangulations}\label{sec2.4}
Having obtained consistent group structure estimates, we proceed to propose a penalized trivariate spline estimation method after grouping for group-specific mean functions $g_k(\cdot)$ based on observations $\left\{Y_{ij}, \boldsymbol{p}_{j}\right\}_{j=1}^{N}$ for $i\in \mathcal{G}_k$. Consider the following trivariate penalized spline regression problem:
\begin{equation}
\min _{s_k \in \mathcal{S}_d^r(\triangle)} \sum_{i\in \mathcal{G}_k}\sum_{j=1}^{N}\left\{Y_{ij}-s_k\left(\boldsymbol{p}_{j}\right)\right\}^2+h_k \mathcal{E}(s_k), \label{objective g}
\end{equation}
where $h_k>0$ is the roughness penalty parameter that controls the smoothness of the fitted spline function $s_k$. Similar to \eqref{smooth object}, we translate \eqref{objective g} to
\begin{equation}
\min_{\boldsymbol{\theta}_k}\sum_{i\in \mathcal{G}_k}\sum_{j=1}^{N}\left\{Y_{ij}-\widetilde{\mathbf{B}}^{\top}\left(\boldsymbol{p}_{j}\right) \boldsymbol{\theta}_k\right\}^2+h_k \boldsymbol{\theta}_k^{\top}\mathbf{D}\boldsymbol{\theta}_k. \label{penalty smooth object}
\end{equation}
Then the minimizer of \eqref{penalty smooth object} is given by 
\begin{equation*}
\widehat{\boldsymbol{\theta}}_k=\left\{\sum_{j=1}^N\widetilde{\mathbf{B}}(\boldsymbol{p}_j)\widetilde{\mathbf{B}}^{\top}(\boldsymbol{p}_j)+\frac{h_k}{|\mathcal{G}_k|}\mathbf{D}\right\}^{-1} \left\{\frac{1}{|\mathcal{G}_k|}\sum_{i\in \mathcal{G}_k}\sum_{j=1}^N\widetilde{\mathbf{B}}(\boldsymbol{p}_j)Y_{ij}\right\}.
\end{equation*}  Consequently, the group mean estimators $\widehat{g}_k(\boldsymbol{p})=\widetilde{\mathbf{B}}^{\top}(\boldsymbol{p}) \widehat{\boldsymbol{\theta}}_k$.

\section{Asymptotic Theory}\label{sec3}
This section establishes the consistency and asymptotic properties of the proposed estimators. To analyze these properties, we introduce some necessary notations and assumptions, which are standard in nonparametric literature; see \cite{huang2003local}, \cite{lai2013bivariate}, \cite{wang2020simultaneous}, \cite{yu2021multivariate}, \cite{li2021sparse}, and \cite{li2024nonparametric}.  

We first introduce some notation. For any function $f$ over the closure of domain $\Omega$, denote the supremum norm as $\|f\|_{\infty, \Omega}=\sup _{\boldsymbol{p} \in \Omega}|f(\boldsymbol{p})|$, and for any $q \geq 0$, denote $|f|_{q, \infty, \Omega}=\max _{|\boldsymbol{\alpha}|=q}\left\|\nabla^{\boldsymbol{\alpha}} f\right\|_{\infty, \Omega}$. Further, we define the standard Sobolev space $\mathcal{W}^{d, \infty}(\Omega)=\left\{f:|f|_{q, \infty, \Omega}<\infty, 0 \leq q \leq d\right\}$. The notation $a\wedge b=\min(a,b)$. Unless otherwise specified, the $c$ and $C$ mentioned in this text uniformly represent positive constants that may change values each time they appear. For two arbitrary sequences,  $a_n$ and $b_n$, depending on $n$, we write $a_n \asymp b_n$ if there exist $c$ and $C$ such that $c|a_n|\leq |b_n|\leq C|a_n|$. 

Next, we introduce some technical assumptions.  
\begin{enumerate}
    \item[(C1)] The true group-specific mean functions $ g_k \in \mathcal{W}^{d+1, \infty}(\Omega)$, $k=1,\ldots,K_0$, for $d \geq 1$.
    
    \item[(C2)] For any $1\leq i\leq n$ and $1\leq j\leq N$, random errors $\varepsilon_{ij}=\varepsilon_{i}\left(\boldsymbol{p}_{j}\right)$ satisfy that $\mathbb{E}\left(\left|\varepsilon_{ij}^{2+\eta}\right|\right) \leq v_{\eta}$, where $\eta$ and $v_{\eta}$ are two positive constants, and $\lim_{h \rightarrow \infty} \mathbb{E}\left\{\varepsilon_{ij}^2 \mathbf{1}(|\varepsilon_{ij}|>h)\right\}=0$, where $\mathbf{1}(\cdot)$ is the indicator function.

    \item[(C3)] The density function of the observations is bounded below and above. 
    %The density function of $\boldsymbol{p}$ is positive continuous and bounded on $\Omega$.

    \item[(C4)] As $N\rightarrow \infty$, the number of the tetrahedra $N_{\tau}$ and the sample size $N$ satisfy that $N_{\tau}\asymp N^\kappa$, where $0<\kappa <1$. The number of the 3D point clouds $n$ and the sample size $N$ satisfy that $n\asymp N^{\nu}$, where $\nu\geq 0$.

    \item[(C5)] The triangulation $\triangle$ is \emph{$\beta$-quasi-uniform}, that is, there exists a positive value $\beta$ such that $\triangle$ satisfies: $|\triangle| / \varrho_\tau \leq \beta<\infty$, for any $\tau \in \triangle$.

    \item[(C6)] Suppose that 
    \begin{align*}
    \min _{1 \leq k \neq k^{\prime} \leq K_0} \min _{i \in \mathcal{G}_k, i^{\prime} \in \mathcal{G}_{k^{\prime}}}\int_{\Omega}|\mu_i(\boldsymbol{p})-\mu_{i^{\prime}}(\boldsymbol{p})|d\boldsymbol{p} =  \min _{1 \leq k \neq k^{\prime} \leq K_0}\int_{\Omega}|g_k(\boldsymbol{p})-g_{k^{\prime}}(\boldsymbol{p})|d\boldsymbol{p} := \omega_{0}>0.
   \end{align*}
    
    \item[(C7)] (i) As $n\rightarrow\infty$, we assume the number of groups $K_0$ to be fixed, and \begin{equation*}
c n \leq \min _{1 \leq k \leq K_0}\left|\mathcal{G}_k\right| \leq \max _{1 \leq k \leq K_0}\left|\mathcal{G}_k\right| \leq Cn.
\end{equation*}
    %In general, for the $k$-th group, we denote $\left|\mathcal{G}_k\right|=n_k$.

    (ii) For any $q$ different (true) groups $\mathcal{G}_{k_1}, \ldots, \mathcal{G}_{k_q}$, when they are falsely merged, define \begin{equation*}
\overline{g}_*(\boldsymbol{p})=\frac{1}{\left|\mathcal{G}_{k_1} \cup \cdots \cup \mathcal{G}_{k_q}\right|}\left\{\left|\mathcal{G}_{k_1}\right| g_{k_1}(\boldsymbol{p})+\cdots+\left|\mathcal{G}_{k_q}\right| g_{k_q}(\boldsymbol{p})\right\} .
\end{equation*}

    There exists $k \in\left\{k_1, \ldots, k_q\right\}$ such that $\int_{\Omega}|\overline{g}_*(\boldsymbol{p})-g_k(\boldsymbol{p})|d\boldsymbol{p}>C$.
    
    (iii) The threshold $\varpi$ satisfies that $\varpi=o(1)$, and $|\triangle|^{d+1}\max_{1\leq k\leq K_0}|g_k|_{d+1, \infty, \Omega}+(\log N)^{1 / 2}N^{-1/2}|\triangle|^{-3/2}=o(\varpi)$.

    \item[(C8)] The penalty parameter satisfies $\max_{1\leq i\leq n}h_i=o\left(N^{1/2} N_{\tau}^{-2/3} \wedge N^{3/4} N_{\tau}^{-1}\right)$. 

\end{enumerate}
\begin{remark}
    Assumptions (C1)--(C5) are commonly used regularity conditions in nonparametric regression and the triangulation-based literature; see \cite{lai2013bivariate}, \cite{yu2019estimation}, \cite{wang2020efficient} and \cite{li2021sparse}. Assumption (C3) is standard in the nonparametric literature. Even though this assumption may not be satisfied with missing data, we can usually find a proper choice of triangulation by adjusting the size of some of the tetrahedra to obtain a decent spline fit. Assumptions (C6) and (C7) are commonly used signal identification conditions in the literature studying latent group structure; see \cite{li2011non}, \cite{chen2019estimating}, and \cite{wang2024panel}. Specifically, Assumption (C6) indicates that the minimum distance of an unknown function between different groups cannot converge to zero, which guarantees the group membership estimate in Theorem \ref{group priori}. Assumption (C7) is used mainly to derive the consistency of the number of groups estimated in Theorem \ref{group estimate}. Assumption (C7)(i) implies that the latent groups have similar sizes (with the same divergence rate). Assumption (C7)(ii) is crucial to prove that $\mathrm{D}(K)$ defined in \eqref{criterion} would be strictly larger than a positive constant when $K<K_0$. The theoretical threshold $\varpi$ in Assumption (C7)(iii) is introduced for proving the consistency of the estimated number of groups. It ensures that $D(K)\to 0$ for $K\ge K_0$, thereby guaranteeing that the ratio criterion can asymptotically separate signal from noise. In practice, $\varpi$ is not necessary; the number of groups can be determined directly by the ratio criterion based on the sequence $D(K)$. Assumption (C8) requires a smaller penalty parameter, which reduces the bias through under-smoothing.
\end{remark}

We start with the asymptotic properties of the estimators obtained without homogeneity pursuit. Theorem \ref{point cloud convergence} presents the global convergence rate for the estimator of the subject mean function in the supremum norm.
\begin{theorem}\label{point cloud convergence}
Under Assumptions (C1)--(C5) and (C8), we have
\begin{equation*}
\max_{1\leq i\leq n}\left\|\widehat{\mu}_{i}-\mu_{i}\right\|_{\infty, \Omega}=O_p\left(|\triangle|^{d+1}\max_{1\leq k\leq K_0}|g_k|_{d+1, \infty, \Omega}+\sqrt{\frac{\log N}{N|\triangle|^{3}}}\right).
\end{equation*}  
\end{theorem}
\begin{remark}
The estimation error rate in Theorem \ref{point cloud convergence} has a standard form as in nonparametric regression; the bias term is of order $O_p\left(|\triangle|^{d+1}\max_{1\leq k\leq K_0}|g_k|_{d+1, \infty, \Omega}\right)$, and the noise term is of order $O_p\left(N^{-1/2}|\triangle|^{-3/2}\sqrt{\log N}\right)$. %This result is a generalization version of Theorem 8 in \cite{li2024nonparametric}.
%Note that, by taking advantage of group structures, the effective sample size becomes $n_kN$, while it is $N$ for the case of full heterogeneity.
\end{remark}

Theorem \ref{group priori} shows the consistency property for the group membership estimate when the number of groups $K_0$ is prespecified.

\begin{theorem}\label{group priori}
Under Assumptions (C1)--(C6) and (C8), suppose that the number of groups $K_0$ is known a priori. Then
\begin{equation}
\mathbb{P}\left(\left\{\widetilde{\mathcal{G}}_k, k=1, \ldots, K_0\right\}=\left\{\mathcal{G}_k, k=1, \ldots, K_0\right\}\right) \rightarrow 1, \quad N\rightarrow \infty.
\label{group priori convergence}
\end{equation}
\end{theorem}
\begin{remark}
By the clustering algorithm, to achieve the consistency property, it is sufficient to show that
\[
\max_{1 \leq i, i' \leq n} \left| \widehat{\Pi}(i, i') - \Pi(i, i') \right|
\]
is of a smaller order than the minimum distance between the true group-specific mean functions. This can be demonstrated by using the global convergence property of the preliminary estimator from Theorem \ref{point cloud convergence}. The consistency result in Theorem \ref{group priori} is similar to some results in the existing literature (although in different model settings), such as \cite{vogt2017classification}, \cite{chen2019estimating}, \cite{vogt2020multiscale}, \cite{chen2021nonparametric}, and \cite{yang2023functional}.
\end{remark}

Theorem \ref{group estimate} shows that, by constructing a ratio criterion \eqref{ratio} in conjunction with Theorem \ref{point cloud convergence}, we can obtain a consistent estimate of the number of groups.

\begin{theorem} \label{group estimate}
Under Assumptions (C1)--(C8), we have
\begin{equation*}
\mathbb{P}\left(\widehat{K}=K_0\right) \rightarrow 1, \quad N\rightarrow \infty.
\end{equation*}
\end{theorem}
\begin{remark} 
The ratio criterion adopted in our approach demonstrates excellent interpretability and ease of implementation, an advantage particularly prominent among comparable methods. In contrast, while the thresholding method proposed by \cite{vogt2020multiscale} and the information criteria developed by \cite{su2019sieve} and \cite{chen2021nonparametric} for selecting the number of groups have both been proven consistent, these approaches fall short of our method in terms of intuitiveness and implementation simplicity. 
\end{remark}

By combining Theorems \ref{group priori} and \ref{group estimate}, we readily obtain the following corollary regarding the consistency of the group structure estimate.
\begin{corollary}\label{coro}
Under Assumptions (C1)--(C8), we have
\begin{equation*}
\mathbb{P}\left(\left\{\widehat{\mathcal{G}}_k, k=1, \ldots, \widehat{K}\right\}=\left\{\mathcal{G}_k, k=1, \ldots, K_0 \right\}\right) \rightarrow 1, \quad N\rightarrow \infty.
\end{equation*}
\end{corollary}
\begin{remark}
This Corollary shows that, with probability approaching one, both the group number and group membership can be estimated correctly by the proposed agglomerative clustering algorithm and ratio criterion. 
\end{remark}

%After obtaining the latent group structure estimation, we can then utilize post-grouping sample information to estimate the group-specific mean functions $\{g_k\}_{k=1}^{K_0}$. 
We further assume the following conditions to derive the consistency and asymptotic normality of the post-grouping estimator. For $k=1,\ldots,K_0$,
\begin{enumerate}
\item[(C4')] The number of tetrahedra $N_{\tau}$ and the sample size $n_kN$ satisfy $N_{\tau}\asymp (n_kN)^\gamma$, where $n_k=|\mathcal{G}_k|$ and $3/(2d+5)<\gamma< 1$.

\item[(C8')] The penalty parameter satisfies $\max_{1\leq k\leq K_0}h_k=o\left(n_k^{1/2}N^{1/2} N_{\tau}^{-2/3} \wedge n_k^{3/4}N^{3/4} N_{\tau}^{-1}\right)$. 
\end{enumerate}

Theorem \ref{group point cloud convergence} shows the global convergence rate of the post-grouping estimator $\widehat{g}_k$ of the group mean function in the supremum norm.
\begin{theorem}\label{group point cloud convergence}
Under Assumptions (C1)--(C4'), (C5), (C7)(i), and (C8'), we have
\begin{align*}
\left\|\widehat{g}_{k}-g_{k}\right\|_{\infty, \Omega}=&O_p\left\{|\triangle|^{d+1}|g_k|_{d+1,\infty,\Omega}+\sqrt{\frac{{\log (n_kN)}}{n_kN|\triangle|^{3}}} \right\}.
%\left\|\widehat{g}_{k}-g_{k}\right\|_{\infty, \Omega}=&O_p\left\{\frac{h_k}{n_kN|\triangle|^{7/2}}|g_k|_{2,\infty,\Omega}+\left(1+\frac{h_k}{n_kN|\triangle|^{11/2}} \right)|\triangle|^{d+1}|g_k|_{d+1,\infty,\Omega}\right.\\
%&\quad\quad \left.+\sqrt{\frac{{\log (n_kN)}}{n_kN|\triangle|^{3}}}+\frac{h_k}{(n_kN)^{3/2}|\triangle|^7} \right\}.
\end{align*}
\end{theorem}
\begin{remark}
This result can drive the optimal convergence rate as shown in \cite{stone1982optimal},
\begin{equation*}
\| \widehat{g}_k^0 -g_k \|_{\infty, \Omega}=O_p\left[\left\{(n_kN)^{-1} \log (n_kN)\right\}^{(d+1) /(2 d+5)}\right],
\end{equation*} 
when $N_{\tau} \asymp \{n_kN/\log (n_kN)\}^{3/(2d+5)}$. 
\end{remark}
Theorem \ref{asymptotic normlity} states the asymptotic normality of the post-grouping estimator $\widehat{g}_k$ at any given point $\boldsymbol{p}\in \Omega$, for $k=1,\ldots, K_0$.
\begin{theorem}\label{asymptotic normlity}
Under Assumptions (C1)--(C4'), (C5), (C7)(i), and (C8'), we have
\begin{equation*} 
\frac{\widehat{g}_{k}(\boldsymbol{p})-g_{k}(\boldsymbol{p})}{\sqrt{\mathrm{Var}\left\{\widehat{g}_{k}(\boldsymbol{p})\right\}}} \stackrel{D} {\longrightarrow} \mathcal{N}\left(0,1\right), \quad n_k\rightarrow \infty,N\rightarrow \infty.
\end{equation*}
\end{theorem}
\begin{remark}
The above asymptotic distribution result can be used to construct asymptotic point-wise confidence intervals in theory. However, it is very difficult to obtain the exact form of the standard error for estimators due to the characteristics of the trivariate spline basis functions. To overcome this, \cite{li2024nonparametric} proposes using a wild bootstrap method to estimate the standard errors of the estimators. After applying the wild bootstrap method to estimate the standard errors, we can further construct confidence intervals for the estimators of group-specific mean functions, thereby quantifying their uncertainty. 
%Suppose that the latent group structure is present, with an unspecified number of groups and unknown membership. In this scenario, we can consistently estimate the group structure by applying Corollary \ref{coro}. Subsequently, we can shift our focus to investigating the asymptotic normality of the estimates for the group-specific functions within each group.
\end{remark}

\section{Simulation Study}\label{sec4}

In this section, we generate 3D point clouds over a horseshoe-shaped domain to mimic complex scenarios encountered in practice. The horseshoe-shaped domain is contained in the rectangular region $[-1, 3.5]\times[-1,1]\times[-0.5,0.5]$, which is discretized by a uniform grid of resolution $129\times54\times30$. We set the total number of 3D point clouds to $n=200$ and $n=500$. In each Monte Carlo replicate, we randomly draw $N=1000$ or $N=2000$ observation locations from this grid, and generate the $n$ point clouds at these locations.

\begin{figure}
\centering
\begin{minipage}{0.24\textwidth}\centering\includegraphics[width=\linewidth]{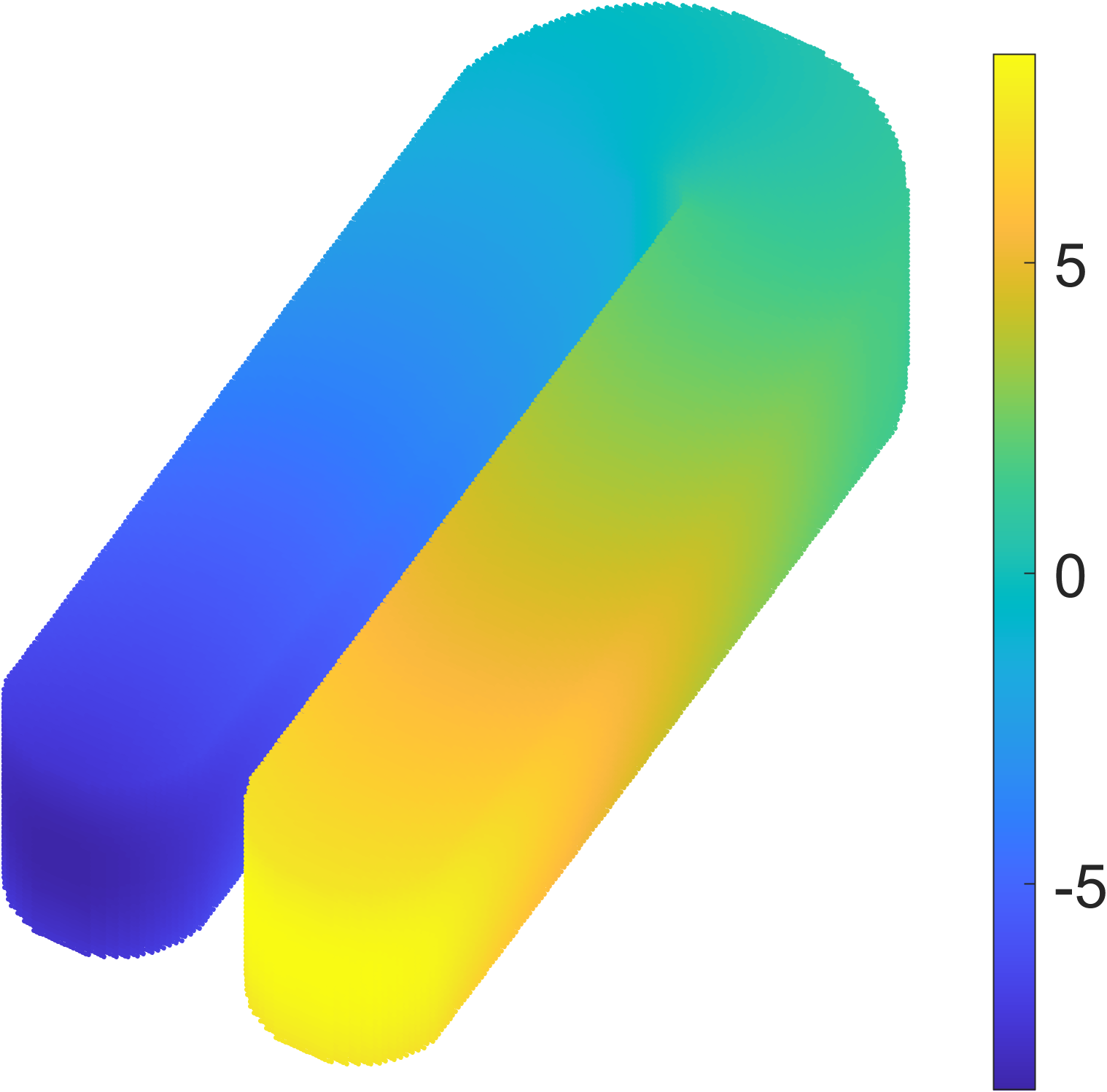}\captionsetup[subfigure]{labelformat=empty}\subcaption{$g_1$}\end{minipage}\hspace{0.05\textwidth}
\begin{minipage}{0.24\textwidth}\centering\includegraphics[width=\linewidth]{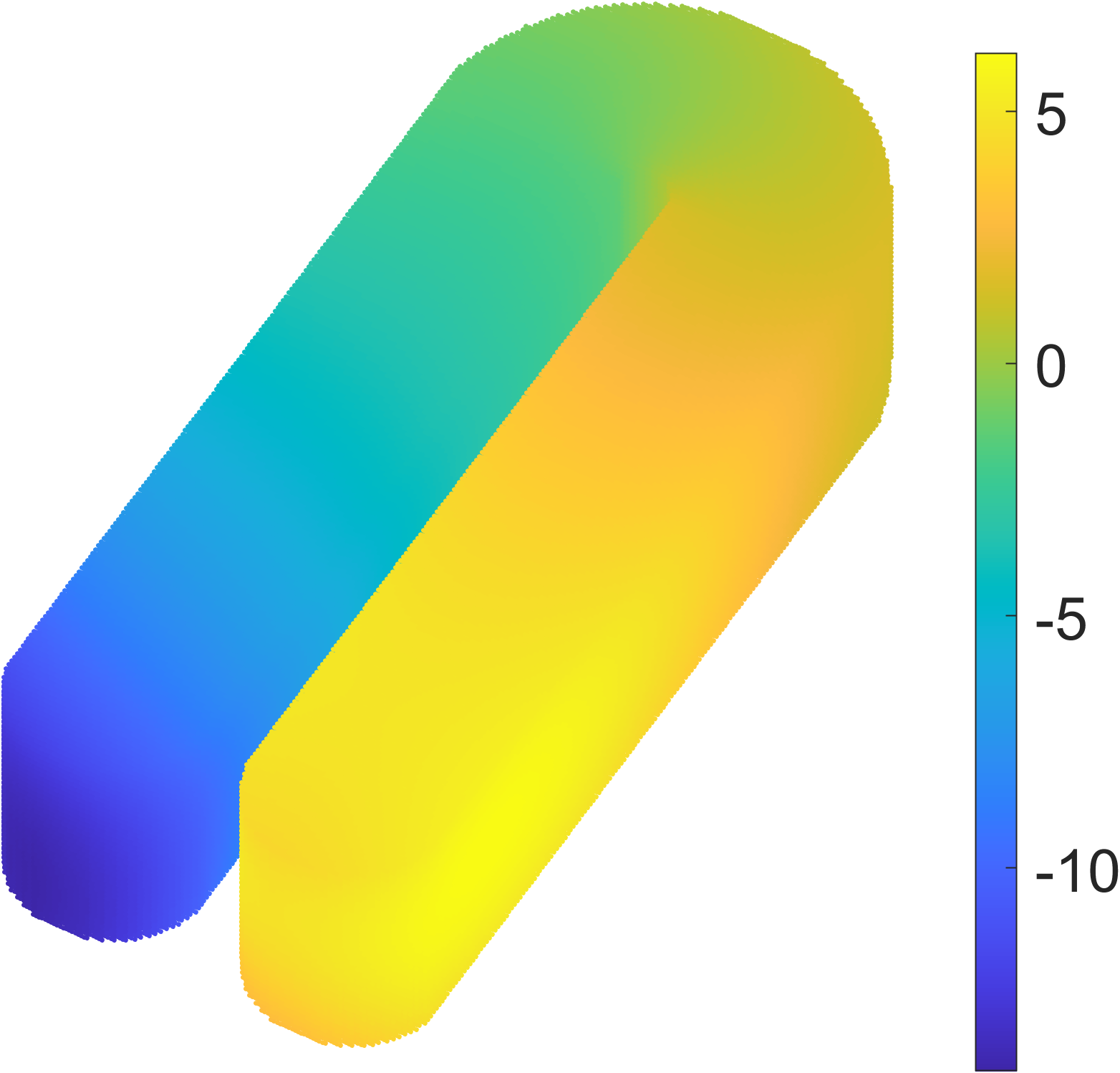}\captionsetup[subfigure]{labelformat=empty}\subcaption{$g_2$}\end{minipage}\hspace{0.05\textwidth}
\begin{minipage}{0.24\textwidth}\centering\includegraphics[width=\linewidth]{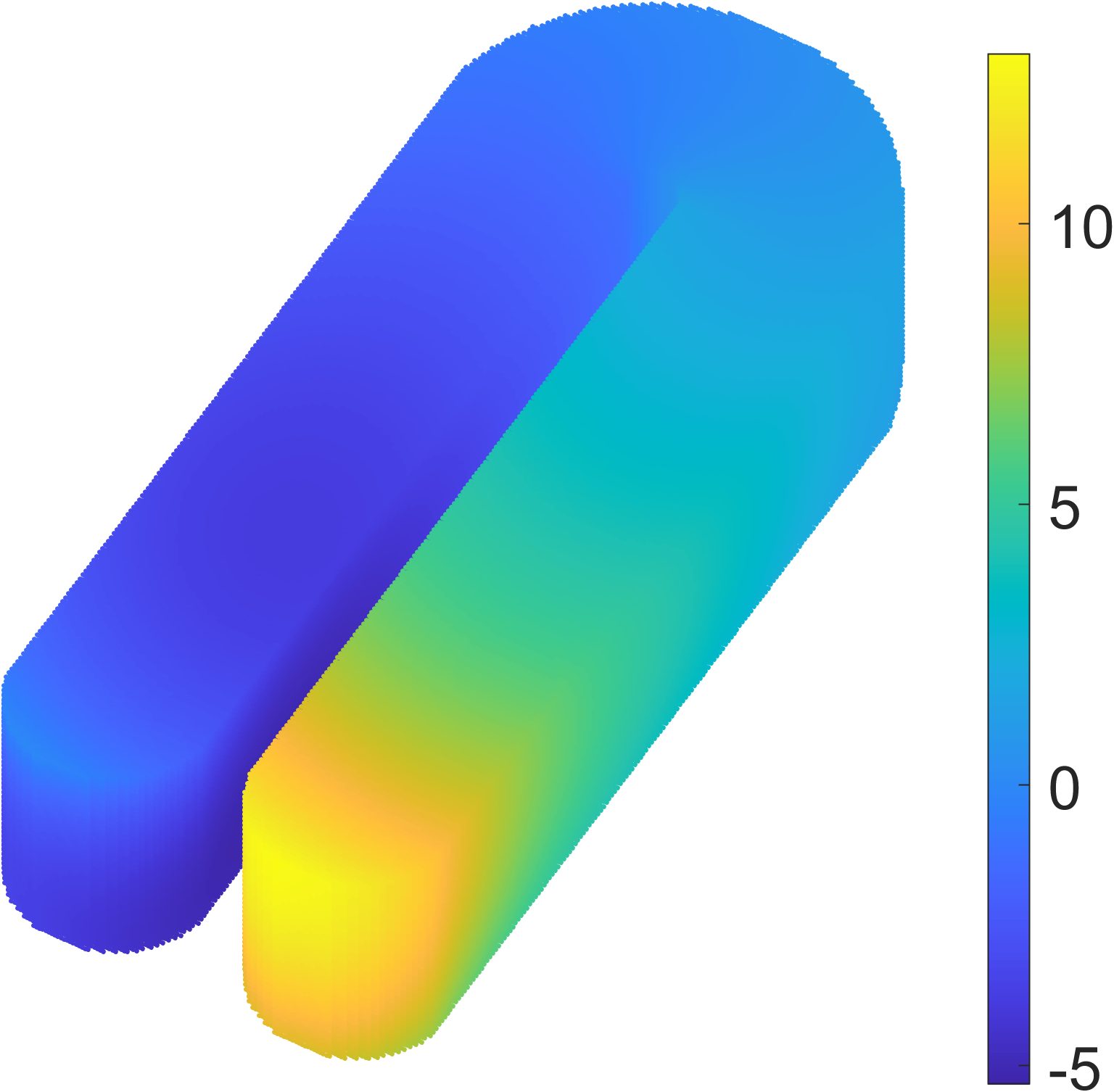}\captionsetup[subfigure]{labelformat=empty}\subcaption{$g_3$}\end{minipage}
\caption{The observed 3D point clouds with different true group-specific mean functions.}\label{simulation1}
\end{figure}

We extract the latent group structure of the underlying signal from the 3D point clouds using models \eqref{model} and \eqref{group}, in which the random noises are drawn from $N(0,1)$.
For the horseshoe-shaped domain, we consider three underlying functions (true group-specific mean functions) with different degrees of variation, $g_1$, $g_2$, and $g_3$, as shown in Figure \ref{simulation1}, so that $K_0=3$. The baseline function is
\begin{equation*}
g_1(x,y,z)=\left\{a(x,y)+d(x,y)^2\right\}(2-z^2),
\end{equation*}
where $a(x,y)$ is the arc-length coordinate along the horseshoe centerline and $d(x,y)$ is the signed distance from the centerline, defined piecewise as
\begin{equation*}
(a,d)=
\begin{cases}
(\pi/4+x, y-0.5), & x\geq 0, y>0,\\
(-\pi/4-x, -0.5-y), & x\ge 0, y\leq 0,\\
\left\{-0.5\arctan(y/x), \sqrt{x^2+y^2}-0.5\right\}, & x<0.
\end{cases}
\end{equation*}
The remaining two functions are defined as
\begin{equation*}
g_k(\boldsymbol{p})=g_1(\boldsymbol{p})+\operatorname{sd}(g_1/2)\,P_k(\boldsymbol{p}),\qquad k=2,3.
\end{equation*}
Here, $\operatorname{sd}(g_1)$ denotes the sample standard deviation of $g_1$ evaluated at the observation locations $\{\boldsymbol{p}_j\}_{j=1}^{N}$, that is,
\begin{equation*}
\operatorname{sd}(g_1)=\left[\frac{1}{N-1}\sum_{j=1}^{N}\left\{g_1(\boldsymbol{p}_j)-\overline{g}_1\right\}^2\right]^{1/2},\qquad 
\overline{g}_1=\frac{1}{N}\sum_{j=1}^{N}g_1(\boldsymbol{p}_j).
\end{equation*}
In addition, $P_k(\boldsymbol{p})$ is a random polynomial in the three coordinates of $\boldsymbol{p}=(x,y,z)^{\top}$ with i.i.d. standard normal coefficients, normalized to have unit sample standard deviation. These group-specific functions $\{g_1,g_2,g_3\}$ generate the 3D point clouds with group sizes $\{0.4n,0.3n,0.3n\}$.

\begin{figure}
\centering
\begin{minipage}{0.2\textwidth}\centering\includegraphics[width=\linewidth]{plot/HS_tri-1.png}\captionsetup[subfigure]{labelformat=empty}\subcaption{$\triangle_1$}\end{minipage}\hspace{0.05\textwidth}
\begin{minipage}{0.2\textwidth}\centering\includegraphics[width=\linewidth]{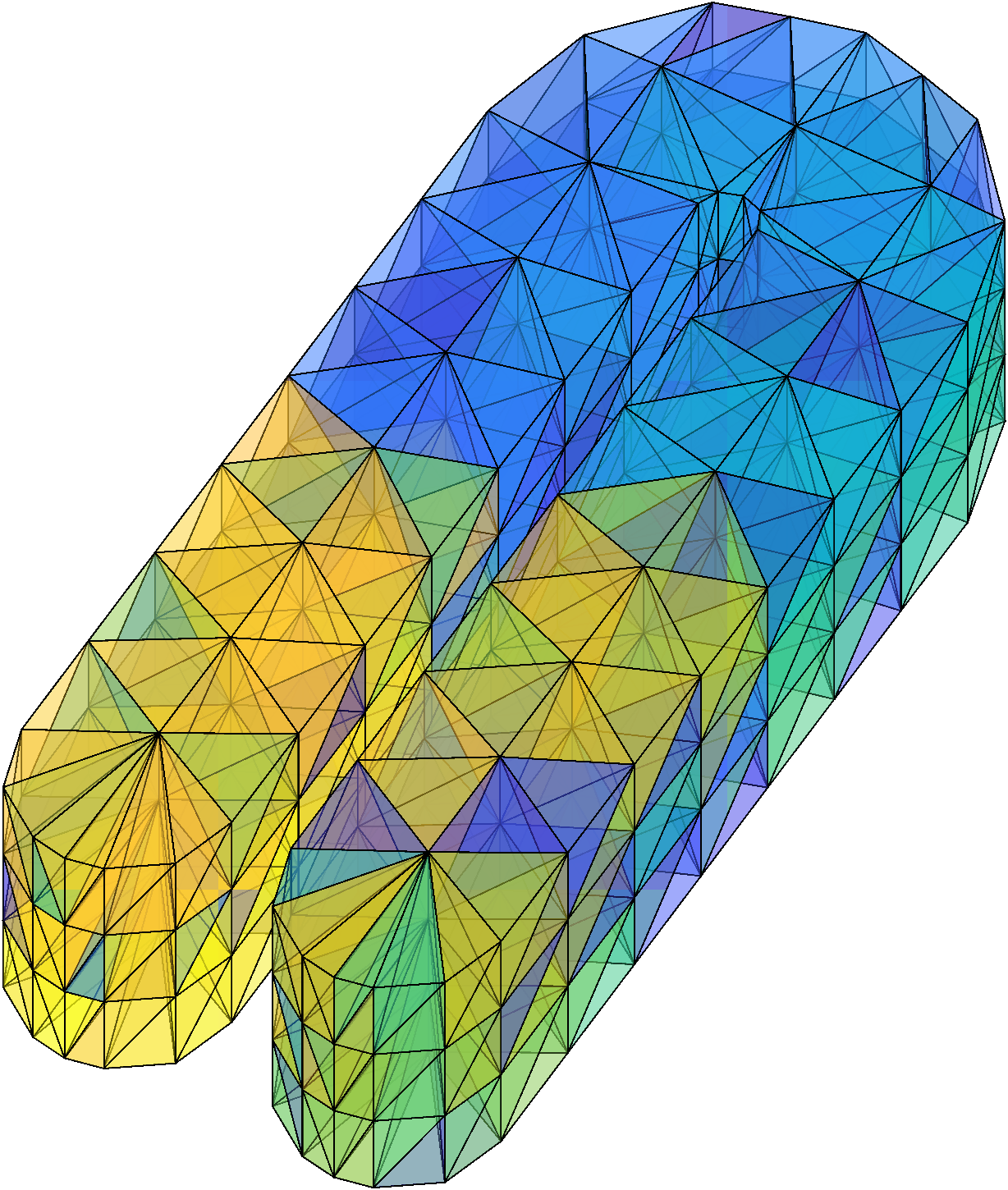}\captionsetup[subfigure]{labelformat=empty}\subcaption{$\triangle_2$}\end{minipage}
\caption{Plot $\triangle_1$ is a relatively coarse triangulation with 504 tetrahedra and 207 vertices, and Plot $\triangle_2$ is a fine triangulation with 756 tetrahedra and 276 vertices.}\label{simulation2}
\end{figure}

Applying the trivariate spline estimation method requires an appropriate triangulation. The construction of the triangulation relies on a geometric domain $\Omega$ defined by polygons for the 3D point clouds. If the domain is unknown, surface reconstruction techniques (e.g., triangular meshes in \cite{shewchuk1998condition}) can be employed to obtain a polygonal domain that tightly encloses the 3D point clouds. An overly coarse triangulation may lead to substantial bias, whereas an excessively fine triangulation can inflate variance and introduce considerable computational burdens. According to Assumption (C4'), the number of tetrahedra $N_\tau$ needs to exceed a minimum value depending on the spline degree. Therefore, the fineness of the triangulation can be determined by setting $N_\tau \approx \lfloor c_1 n N^\gamma \log(nN)\rfloor + c_2$, where $\lfloor\cdot\rfloor$ denotes the floor function, $c_1$ and $c_2$ are tuning constants, and $3/(2d+5)<\gamma<1$. Once the triangulation is sufficiently fine, further refinement typically has negligible impact on fitting accuracy. Figure \ref{simulation2} displays two triangulations for the 3D point clouds over the horseshoe-shaped domain. For the spline smoothing parameters, we set the smoothness order $r=1$ and consider polynomial degrees $d=2,3$ and $4$. Smaller values of $d$ yield smoother and more parsimonious estimates, whereas larger values provide additional local flexibility. This choice balances bias, variance, and computational cost. The penalty parameter is selected by the generalized cross-validation (GCV) criterion (\cite{craven1978smoothing}) on an 11-point grid over which $\log_{10}(h)$ is uniformly spaced on $[-5,0]$. For the proposed method, the number of groups is estimated by the ratio criterion \eqref{ratio} with an upper bound $\overline{K}=5$.

We benchmark the proposed method, denoted TPST-AC (TPST with agglomerative clustering), against three clustering approaches. The first is the classical K-means (KM), which directly applies the KM algorithm (\cite{macqueen1967some}) to the observed point-cloud vectors. The second is principal component analysis followed by KM (PCA-KM), in which the leading principal components (\cite{jolliffe2002principal}) are first extracted and then clustered by KM. The third is kernel ridge regression followed by KM (KRR-KM), in which each point cloud is first smoothed by kernel ridge regression (\cite{saunders1998ridge}) and the resulting fitted curves are then clustered by KM. For the comparison methods, the number of groups is selected using a criterion natural to each method. For KM and PCA-KM, which do not possess an explicit likelihood, the number of groups is chosen by maximizing the average silhouette coefficient \cite{rousseeuw1987silhouettes}. For KRR-KM, the number of groups is chosen by minimizing a generalized information criterion \cite{chen2019estimating}, constructed from the pooled residual sum of squares of the kernel ridge estimates.

We evaluate the performance of latent group structure estimation by calculating the empirical probability of correctly recovering the number of groups under different scenarios. The empirical probability is calculated as follows:
\begin{equation*}
\operatorname{EP}(\widehat{K}=K_0)=\frac{1}{M}\sum_{m=1}^{M}\mathbf{1}\{\widehat{K}^{(m)}=K_0\},
\end{equation*}
based on $M=200$ Monte Carlo replications. We further assess the accuracy of the estimated groups and group memberships produced by the agglomerative grouping algorithm in Section \ref{sec2.2}. We consider the purity between the estimated groups $\widehat{\mathcal{G}}=\{\widehat{\mathcal{G}}_1,\ldots,\widehat{\mathcal{G}}_{\widehat{K}}\}$ and the true groups $\mathcal{G}=\{\mathcal{G}_1,\ldots,\mathcal{G}_{K_0}\}$, a classic criterion of grouping quality defined as
\begin{equation*}
\operatorname{Purity}(\widehat{\mathcal{G}}, \mathcal{G})=\frac{1}{n} \sum_{k=1}^{\widehat{K}} \max_{1 \leq l \leq K_0}\left|\widehat{\mathcal{G}}_k \cap \mathcal{G}_l\right|.
\end{equation*}
We further report the normalized mutual information (NMI),
\begin{equation*}
\operatorname{NMI}(\widehat{\mathcal{G}},\mathcal{G})=\frac{2I(\widehat{\mathcal{G}},\mathcal{G})}{H(\widehat{\mathcal{G}})+H(\mathcal{G})},
\end{equation*}
where $I$ denotes the mutual information and $H$ denotes the entropy.
We also compute the adjusted Rand index (ARI),
\begin{equation*}
\operatorname{ARI}=\frac{\sum_{kl}\binom{n_{kl}}{2}-\left[\sum_k\binom{a_k}{2}\sum_l\binom{b_l}{2}\right]/\binom{n}{2}}
{\frac12\left[\sum_k\binom{a_k}{2}+\sum_l\binom{b_l}{2}\right]-\left[\sum_k\binom{a_k}{2}\sum_l\binom{b_l}{2}\right]/\binom{n}{2}},
\end{equation*}
with $n_{kl}=|\widehat{\mathcal{G}}_k\cap\mathcal{G}_l|$, $a_k=\sum_l n_{kl}$, and $b_l=\sum_k n_{kl}$. The closer the purity, NMI, and ARI values are to $1$, the more accurately the estimated groups correspond to the true ones.

We also examine the effect of grouping by evaluating the accuracy of two estimation methods, namely the preliminary penalized estimators $\{\widehat{\mu}_i\}_{i=1}^n$ defined in Section \ref{sec2.1} and the post-grouping estimators $\{\widehat{g}_k\}_{k=1}^{\widehat{K}}$ defined in Section \ref{sec2.4}. We use the average root mean square error (RMSE) to evaluate each estimation method. The average RMSE for the preliminary penalized estimator is defined as
\begin{equation*}
\operatorname{RMSE}(\widehat{\mu})=\frac{1}{n}\sum_{i=1}^n \left[\frac{1}{N}\sum_{j=1}^N \left(\widehat{\mu}_i(\boldsymbol{p}_j)-\mu_{i}(\boldsymbol{p}_j) \right)^2 \right]^{1/2}.
\end{equation*}
The average RMSE of the post-grouping estimator is defined as
\begin{equation*}
\operatorname{RMSE}(\widehat{g})=\frac{1}{n}\sum_{k=1}^{\widehat{K}}|\widehat{\mathcal{G}}_k| \left[\frac{1}{N}\sum_{j=1}^N \left(\widehat{g}_k(\boldsymbol{p}_j)-g_{k}(\boldsymbol{p}_j) \right)^2 \right]^{1/2}.
\end{equation*}
For KRR-KM, which also produces preliminary and pooled estimates, the two RMSE measures are computed with the same definitions as for TPST-AC. In contrast, PCA-KM and KM only assign group memberships and do not estimate the underlying mean functions. We also report the average computation time of each method to compare computational costs.

\begin{table}
  \centering
  \caption{Comparison of grouping recovery, function estimation, and computational cost of different methods for $n=200$.}
  \label{tab:a} 
  \resizebox{\textwidth}{!}{%
  \begin{tabular}{@{}ccccccccccc@{}}
    \toprule
    $N$ & Method & Triangulation & $d$ & Purity & NMI & ARI & EP($\widehat{K}=K_0$) & RMSE($\widehat{\mu}$) & RMSE($\widehat{g}$) & Time(s) \\
    \midrule
    
    1000 & TPST-AC & $\triangle_1$ & 2 & 0.9760 & 0.9811 & 0.9714 & 92.00\% & 0.2724 & 0.1985 & 0.0288 \\
    \cmidrule(lr){2-11}
    & TPST-AC & $\triangle_1$ & 3 & 0.9760 & 0.9811 & 0.9714 & 92.00\% & 0.2309 & 0.0530 & 0.1692 \\
    \cmidrule(lr){2-11}
    & TPST-AC & $\triangle_1$ & 4 & 0.9790 & 0.9827 & 0.9735 & 92.50\% & 0.2441 & 0.0575 & 5.0490 \\
    \cmidrule(lr){2-11}
    & TPST-AC & $\triangle_2$ & 2 & 0.9745 & 0.9799 & 0.9696 & 91.50\% & 0.2724 & 0.1986 & 0.0297 \\
    \cmidrule(lr){2-11}
    & TPST-AC & $\triangle_2$ & 3 & 0.9730 & 0.9788 & 0.9678 & 91.00\% & 0.2282 & 0.0539 & 0.1700 \\
    \cmidrule(lr){2-11}
    & TPST-AC & $\triangle_2$ & 4 & 0.9890 & 0.9894 & 0.9854 & 92.50\% & 0.2483 & 0.0548 & 10.178 \\
    \cmidrule(lr){2-11}
    & KRR-KM  & -- & -- & 0.9959 & 0.9840 & 0.9782 & 83.50\% & 0.5448 & 0.1130 & 5.2191 \\
    \cmidrule(lr){2-11}
    & PCA-KM & -- & --  & 0.9670 & 0.9740 & 0.9607 & 89.00\% & -- & -- & 0.0237 \\
    \cmidrule(lr){2-11}
    & KM & -- & --  & 0.7555 & 0.7858 & 0.6600 & 18.00\% & -- & -- & 0.0439 \\
    \midrule
    
    2000 & TPST-AC & $\triangle_1$ & 2 & 0.9834 & 0.9866 & 0.9801 & 94.50\% & 0.2433 & 0.1957 & 0.0388 \\
    \cmidrule(lr){2-11}
    & TPST-AC & $\triangle_1$ & 3 & 0.9862 & 0.9883 & 0.9828 & 95.00\% & 0.1820 & 0.0408 & 0.1919 \\
    \cmidrule(lr){2-11}
    & TPST-AC & $\triangle_1$ & 4 & 0.9775 & 0.9823 & 0.9732 & 92.50\% & 0.1825 & 0.0465 & 6.2736 \\
    \cmidrule(lr){2-11}
    & TPST-AC & $\triangle_2$ & 2 & 0.9835 & 0.9869 & 0.9803 & 94.50\% & 0.2432 & 0.1957 & 0.0372 \\
    \cmidrule(lr){2-11}
    & TPST-AC & $\triangle_2$ & 3 & 0.9814 & 0.9848 & 0.9773 & 94.00\% & 0.1788 & 0.0420 & 0.1875 \\
    \cmidrule(lr){2-11}
    & TPST-AC & $\triangle_2$ & 4 & 0.9832 & 0.9863 & 0.9797 & 94.50\% & 0.1879 & 0.0474 & 12.062 \\
    \cmidrule(lr){2-11}
    & KRR-KM & -- & --  & 1.0000 & 0.9980 & 0.9980 & 96.00\% & 0.4727 & 0.1042 & 37.694 \\
    \cmidrule(lr){2-11}
    & PCA-KM & -- & --  & 0.9790 & 0.9835 & 0.9750 & 93.00\% & -- & -- & 0.0440 \\
    \cmidrule(lr){2-11}
    & KM & -- & --  & 0.7585 & 0.7883 & 0.6647 & 18.00\% & -- & -- & 0.1296 \\
    \bottomrule
  \end{tabular}%
  }
\end{table}

\begin{table}
  \centering
  \caption{Comparison of grouping recovery, function estimation, and computational cost of different methods for $n=500$.}
  \label{tab:b} 
  \resizebox{\textwidth}{!}{%
  \begin{tabular}{@{}ccccccccccc@{}}
    \toprule
    $N$ & Method & Triangulation & $d$ & Purity & NMI & ARI & EP($\widehat{K}=K_0$) & RMSE($\widehat{\mu}$) & RMSE($\widehat{g}$) & Time(s) \\
    \midrule
    1000 & TPST-AC & $\triangle_1$ & 2 & 0.9775 & 0.9821 & 0.9728 & 92.50\% & 0.2727 & 0.1945 & 0.0659 \\
    \cmidrule(lr){2-11}
    & TPST-AC & $\triangle_1$ & 3 & 0.9760 & 0.9803 & 0.9704 & 91.00\% & 0.2312 & 0.0404 & 0.2546 \\
    \cmidrule(lr){2-11}
    & TPST-AC & $\triangle_1$ & 4 & 0.9805 & 0.9835 & 0.9752 & 92.50\% & 0.2444 & 0.0440 & 6.2858 \\
    \cmidrule(lr){2-11}
    & TPST-AC & $\triangle_2$ & 2 & 0.9775 & 0.9821 & 0.9728 & 92.50\% & 0.2726 & 0.1943 & 0.0605 \\
    \cmidrule(lr){2-11}
    & TPST-AC & $\triangle_2$ & 3 & 0.9760 & 0.9809 & 0.9710 & 92.00\% & 0.2285 & 0.0408 & 0.2593 \\
    \cmidrule(lr){2-11}
    & TPST-AC & $\triangle_2$ & 4 & 0.9834 & 0.9861 & 0.9795 & 90.00\% & 0.2486 & 0.0438 & 11.094 \\
    \cmidrule(lr){2-11}
    & KRR-KM  & -- & -- & 0.9985 & 0.9679 & 0.9536 & 72.00\% & 0.5449 & 0.0776 & 5.4977 \\
    \cmidrule(lr){2-11}
    & PCA-KM & -- & --  & 0.9640 & 0.9717 & 0.9572 & 88.00\% & -- & -- & 0.0707 \\
    \cmidrule(lr){2-11}
    & KM & -- & --  & 0.7480 & 0.7775 & 0.6468 & 15.50\% & -- & -- & 0.1462 \\
    \midrule
    
    2000 & TPST-AC & $\triangle_1$ & 2 & 0.9925 & 0.9941 & 0.9911 & 97.50\% & 0.2445 & 0.1931 & 0.0982 \\
    \cmidrule(lr){2-11}
    & TPST-AC & $\triangle_1$ & 3 & 0.9924 & 0.9938 & 0.9909 & 97.50\% & 0.1821 & 0.0294 & 0.3892 \\
    \cmidrule(lr){2-11}
    & TPST-AC & $\triangle_1$ & 4 & 0.9880 & 0.9905 & 0.9857 & 96.00\% & 0.1827 & 0.0340 & 7.7504 \\
    \cmidrule(lr){2-11}
    & TPST-AC & $\triangle_2$ & 2 & 0.9940 & 0.9948 & 0.9923 & 97.50\% & 0.2444 & 0.1930 & 0.1035 \\
    \cmidrule(lr){2-11}
    & TPST-AC & $\triangle_2$ & 3 & 0.9895 & 0.9917 & 0.9875 & 96.50\% & 0.1790 & 0.0293 & 0.3077 \\
    \cmidrule(lr){2-11}
    & TPST-AC & $\triangle_2$ & 4 & 0.9893 & 0.9910 & 0.9870 & 96.00\% & 0.1880 & 0.0357 & 14.483 \\
    \cmidrule(lr){2-11}
    & KRR-KM & -- & --  & 1.0000 & 0.9875 & 0.9821 & 88.50\% & 0.4740 & 0.0718 & 40.949 \\
    \cmidrule(lr){2-11}
    & PCA-KM & -- & --  & 0.9820 & 0.9858 & 0.9786 & 94.00\% & -- & -- & 0.1086 \\
    \cmidrule(lr){2-11}
    & KM & -- & --  & 0.7495 & 0.7784 & 0.6480 & 16.00\% & -- & -- & 0.2799 \\
    \bottomrule
  \end{tabular}%
  }
\end{table}

Tables \ref{tab:a} and \ref{tab:b} summarize the grouping and estimation results for $n=200$ and $n=500$. For TPST-AC, the empirical probability $\operatorname{EP}(\widehat{K}=K_0)$ is mainly determined by the number of observation locations, $N$. As $N$ increases from $1000$ to $2000$, this probability also increases, whereas the number of point clouds, $n$, plays a relatively minor role, although a larger $n$ can further improve recovery in some settings. This pattern is consistent with the construction of the ratio criterion; the ability of this criterion to recover the group structure depends mainly on the accuracy of the individual point-cloud estimates, and this accuracy improves as $N$ increases. The two triangulations produce nearly identical grouping results, indicating that once the triangulation is sufficiently fine, further refinement has little effect on group identification. In particular, when $n=500$, $N=1000$, and $d=2$, TPST-AC under $\triangle_1$ and $\triangle_2$ yields identical values for all grouping criteria. The spline degree, $d$, likewise has little effect on group identification: purity, NMI, ARI, and EP remain comparable across spline degrees, and for $n=200$, $N=1000$, and $\triangle_1$, TPST-AC with $d=2$ and $d=3$ again yields identical values. This is because the accuracy of the initial estimate, as reflected by $\operatorname{RMSE}(\widehat{\mu})$, differs only slightly across the different spline degrees. However, spline degree $d$ does affect the accuracy of the post-grouping estimate. Relative to $d=2$, using $d=3$ or $d=4$ substantially reduces the RMSE of the post-grouping estimate $\widehat{g}$.

Among the competing methods, raw KM performs the worst. It clusters the noisy point-cloud vectors directly, without any smoothing or dimension reduction, and thus is fundamentally unable to recover the latent group structure; this is reflected in the fact that its purity, NMI, ARI, and EP are the lowest among all methods. PCA-KM achieves substantial improvement by clustering the leading principal components, and its grouping indices are also competitive. However, it remains a membership-only procedure: it does not estimate the group-specific mean functions, and its ability to recover the correct number of groups is generally inferior to that of TPST-AC. In terms of grouping accuracy, KRR-KM is the strongest competitor and can attain the highest purity in some settings. This apparent strength, however, stems from KRR-KM's tendency to split a true group into smaller but still pure subgroups, whereas purity does not penalize over-splitting. This can lead to an overestimated number of groups and unstable EP, so its EP is often lower than that of TPST-AC. In addition, KRR-KM produces less accurate preliminary estimates; its post-grouping RMSE is generally larger than that of TPST-AC, and its computation time is substantially longer. By contrast, TPST-AC simultaneously provides reliable group identification and accurate mean-function estimation, without suffering from the computational burden or the over-splitting tendency of the competing methods. Overall, among the methods considered, TPST-AC offers the most balanced and practically attractive solution for clustering and estimating 3D point clouds.

\section{ADNI Data Analysis}\label{sec5}

We illustrate the application of our proposed method using the MRI data from ADNI, which is publicly available and accessible through \url{http://adni.loni.usc.edu/}. The database we used consists of brain MRIs from 937 diagnosed subjects, including 120 with AD, 667 with mild cognitive impairment (MCI), and 150 cognitive normals (CN). MRI, as a non-invasive imaging technique, plays an indispensable role in diagnosing and researching AD. It provides high-resolution images of brain structure, especially in the early stages of the disease when cognitive symptoms are not yet apparent, making MRI a key tool for identifying and monitoring AD. 

\begin{figure}
\centering
\begin{minipage}{0.27\textwidth}\centering\includegraphics[width=\linewidth]{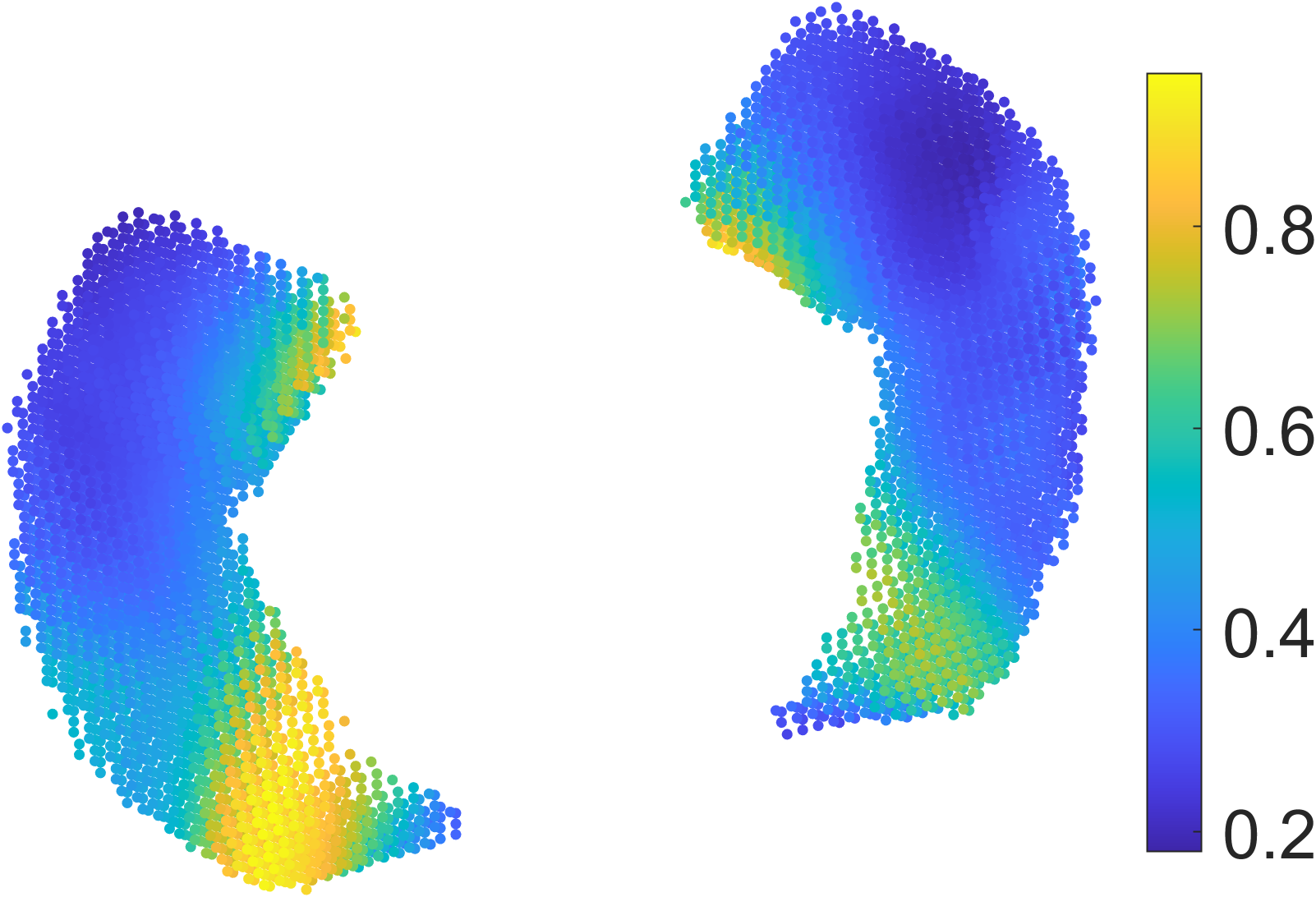}\captionsetup[subfigure]{labelformat=empty}\subcaption{AD}\end{minipage}\hspace{0.05\textwidth}
\begin{minipage}{0.27\textwidth}\centering\includegraphics[width=\linewidth]{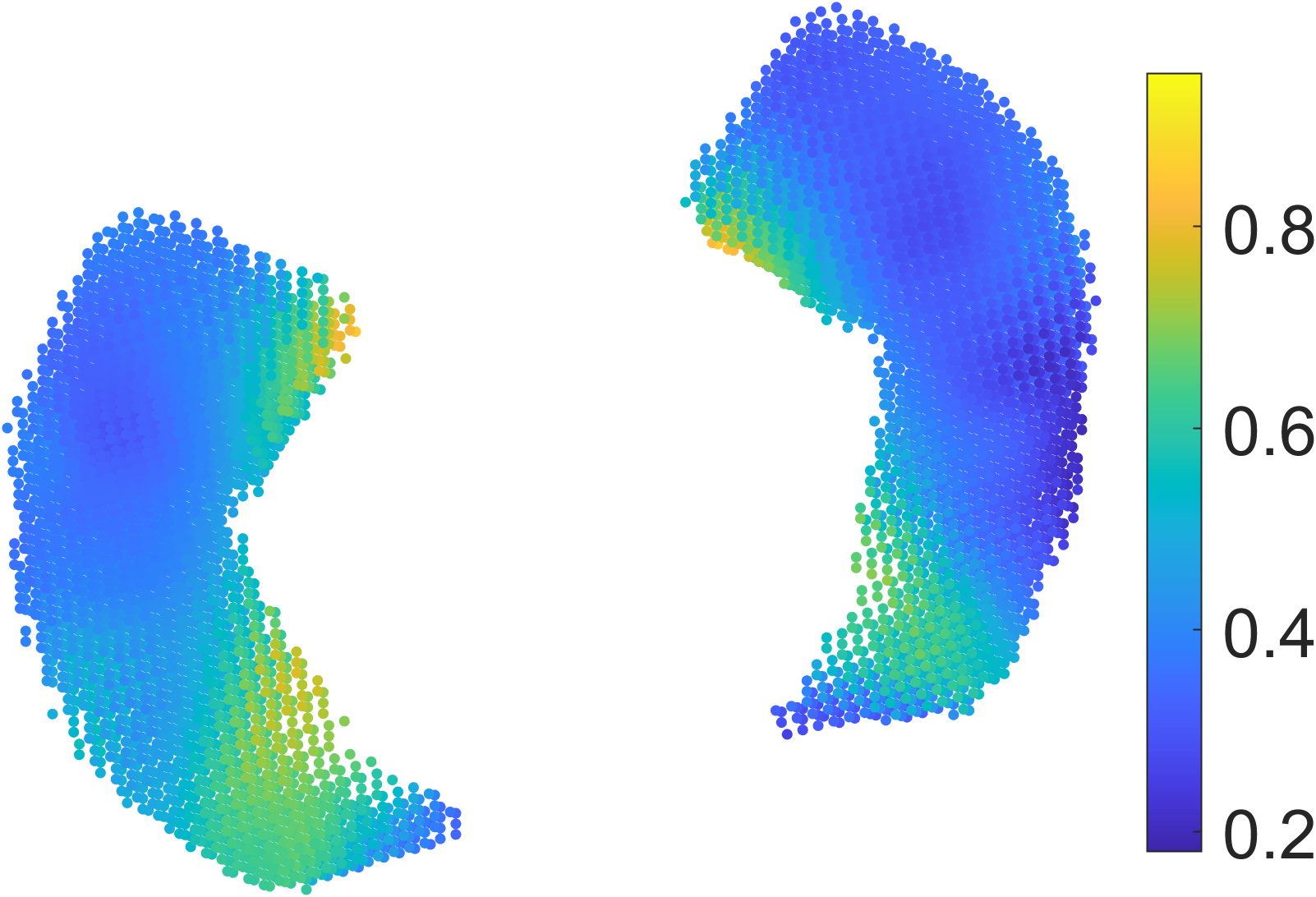}\captionsetup[subfigure]{labelformat=empty}\subcaption{MCI}\end{minipage}\hspace{0.05\textwidth}
\begin{minipage}{0.27\textwidth}\centering\includegraphics[width=\linewidth]{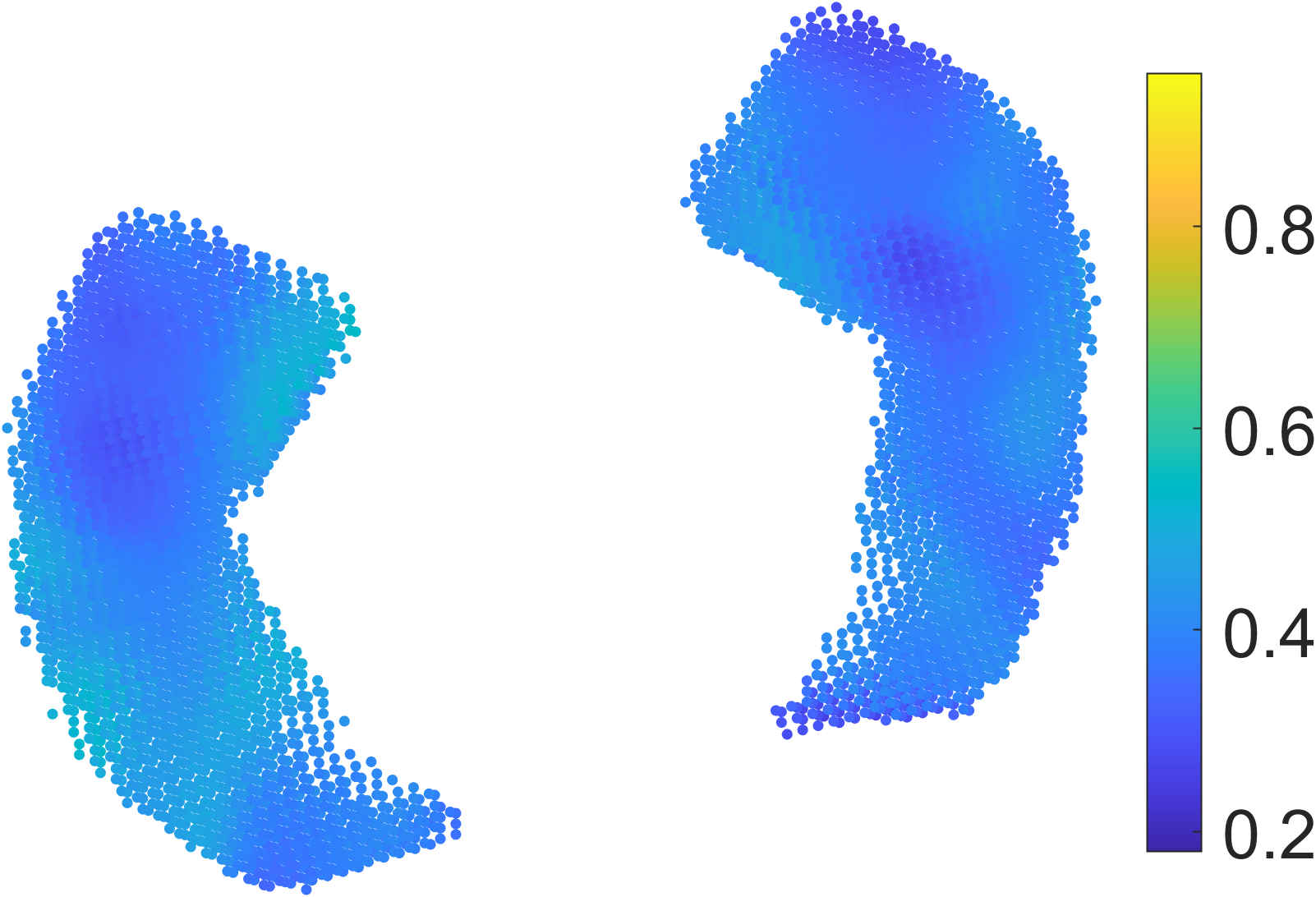}\captionsetup[subfigure]{labelformat=empty}\subcaption{CN}\end{minipage}
\caption{Comparison of left (3,447 voxels) and right (3,608 voxels) hippocampus images from subjects across the three diagnostic statuses.}\label{hippocampus}
\end{figure}

3D MRI data contains many voxels, and this vast scale of data undoubtedly poses a serious challenge to computational capabilities and memory constraints. Moreover, not all signals within the brain are associated with AD, and numerous low-value data points can significantly affect the efficiency and accuracy of group identification. Existing medical research identifies the hippocampus as one of the core regions affected in early AD. Its atrophy and dysfunction are closely linked to memory impairment, and its structural changes are often among the earliest pathological features; see \cite{van1990hippocampal} and \cite{rao2022hippocampus}. Therefore, we focused on the hippocampus, aiming to identify and assess the progression of AD lesions more efficiently through in-depth analysis of this key area. Each hippocampus was extracted in standard space using the FreeSurfer hippocampal labels, and the left and right regions were combined into a common voxel grid so that all subjects share the same set of observation locations. The intensity values were then normalized within each hemisphere by a min--max transformation, which preserves the AD/MCI/CN differences in mean intensity level while removing scale differences across scans. Figure \ref{hippocampus} displays the structure of the hippocampus on the left and right sides of some subjects, with each dataset containing 7,055 voxels. We can observe that the hippocampal MRI data in the AD and MCI groups exhibit significantly abnormal high-signal characteristics when compared to the CN group.

\begin{figure}
\centering
\includegraphics[width=0.25\linewidth]{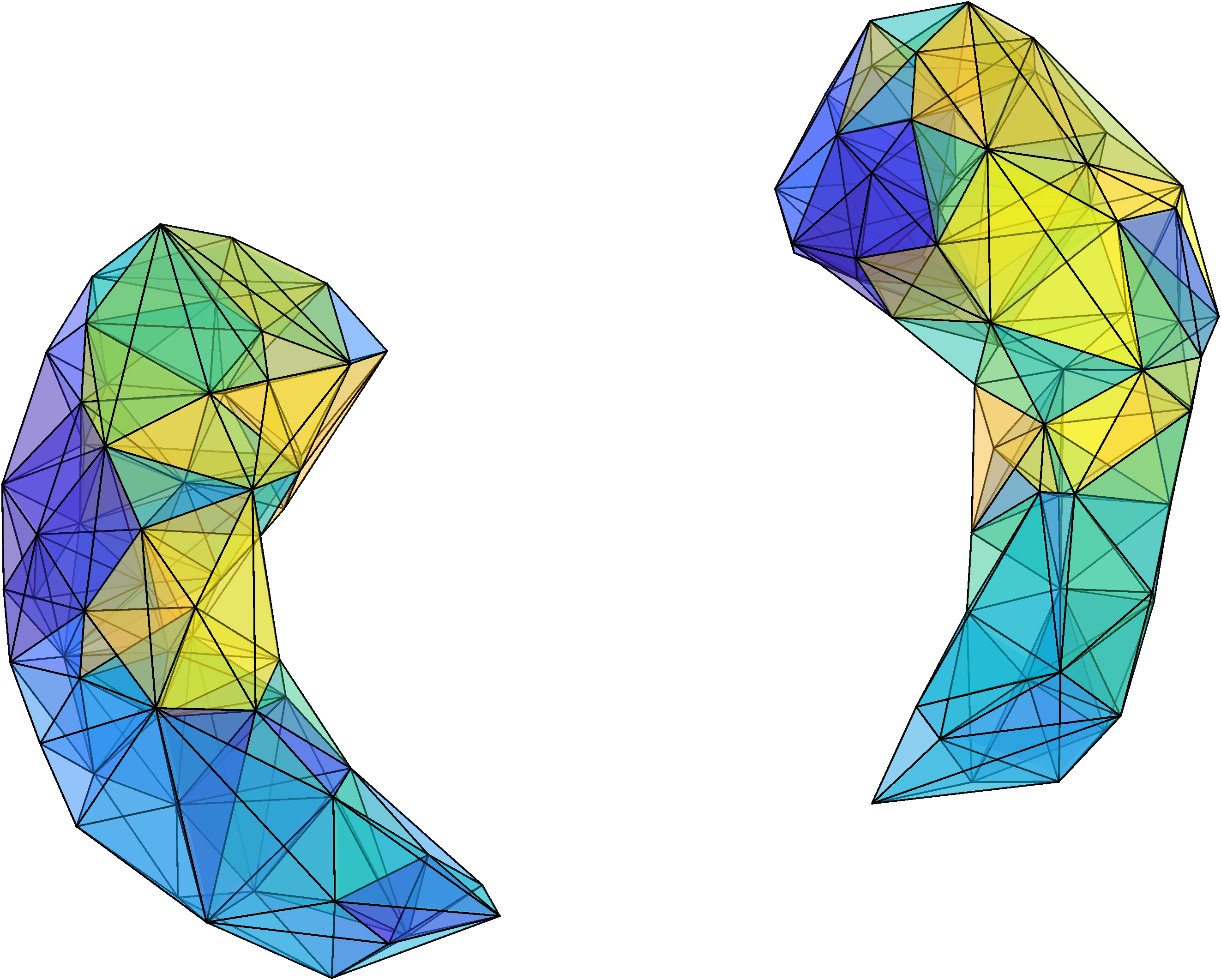} 
\caption{Triangulation of the left (145 tetrahedra, 59 vertices) and right (151 tetrahedra, 62 vertices) hippocampus.}\label{hippocampus tri}
\end{figure}

We apply the proposed TPST-AC method to the MRI hippocampus data and compare its performance with KM, PCA-KM, and KRR-KM. To effectively capture the high-frequency oscillations in the human brain scans, we set the degree $d=4$ and the smoothness parameter $r=1$ to generate trivariate splines. Figure \ref{hippocampus tri} illustrates the triangulation of the hippocampus structure on both the left and right sides, which contains 296 tetrahedra and 121 vertices. We set $\overline{K}=5$ as the upper bound for the ratio criterion \eqref{ratio}.

We treat the clinical diagnostic categories as the true group labels. Table \ref{tab:adni_methods} reports the estimated number of groups and the clustering accuracy for the four methods. TPST-AC selects three groups, and the resulting partition is highly consistent with the diagnostic labels. By contrast, KM identifies only two groups and yields low clustering metrics, indicating that directly clustering the raw hippocampal intensity vectors is insufficient to recover the AD, MCI, and CN structure. PCA-KM gives the same estimated number of groups and clustering accuracy as KM, suggesting that extracting principal components before clustering does not improve the grouping on this real data. KRR-KM attains a high purity but selects five groups, with markedly lower NMI and ARI; it therefore tends to over-split the true groups, consistent with the simulation findings.

\begin{table}
\centering
\caption{Comparison of grouping results for the ADNI hippocampus MRI data across methods.}
\label{tab:adni_methods}
\begin{tabular}{ccccc}
\toprule
Method & $\widehat K$ & Purity & NMI & ARI \\
\midrule
TPST-AC & 3 & 0.9829 & 0.9000 & 0.9408 \\
KM & 2 & 0.7118 & 0.2337 & 0.0809 \\
PCA-KM & 2 & 0.7118 & 0.2337 & 0.0809 \\
KRR-KM & 5 & 0.9349 & 0.4719 & 0.2804 \\
\bottomrule
\end{tabular}
\end{table}

Table \ref{tab:adni_groups} reports the composition of the three groups estimated by TPST-AC. Two groups consist entirely of AD and CN subjects, respectively, while the remaining group is dominated by MCI subjects and also contains a small number of AD and CN subjects. This pattern is consistent with the continuous progression of hippocampal atrophy in the disease course. There is no sharp boundary between AD and MCI, so some subjects are in a transitional state in terms of imaging features; normal aging can also induce hippocampal atrophy, making a few cognitively normal subjects overlap with early-stage MCI. These boundary subjects are therefore assigned to the MCI-dominated intermediate group, which does not indicate a failure of clustering. Rather, it suggests that the proposed method can capture continuous variation beyond the diagnostic labels. This provides a useful clue for using hippocampal atrophy patterns to assist early identification of AD and staging of disease progression.

\begin{table}
\centering
\caption{Composition of the three groups recovered by TPST-AC.}
\label{tab:adni_groups}
\begin{tabular}{cc}
\toprule
Clustering & Main Composition (Number) \\
\midrule
Group 1 & AD (114) \\
Group 2 & MCI (667) + CN (10) + AD (6) \\
Group 3 & CN (140) \\
\bottomrule
\end{tabular}
\end{table}

Overall, TPST-AC recovers a group structure that is highly consistent with the clinically assigned diagnostic status, and it shows clear advantages over the competing methods in both the estimated number of groups and clustering accuracy. These results indicate that the proposed method can effectively identify latent group structure in noisy, irregular real three-dimensional point clouds.

\section{Conclusion}\label{sec6}

This paper proposes a classification learning method for identifying the latent group structure of 3D point clouds in irregular domains. The method first obtains smooth subject-specific mean function estimators by penalized trivariate splines, then determines both the number of groups and their membership through a ratio criterion with agglomerative clustering, and finally pools the point clouds within each estimated group to produce group-specific mean function estimators. Applied to the ADNI hippocampus MRI data, the method recovers three groups that correspond closely to the AD, MCI, and CN diagnoses, supporting the role of hippocampal atrophy as a core marker of AD and its early stages (\cite{van1990hippocampal} and \cite{rao2022hippocampus}). The MCI-dominated group also absorbs a small number of boundary AD and CN subjects, reflecting the continuous progression of hippocampal atrophy and the overlap between early pathological changes and normal aging. This suggests that clinical group structure is often better viewed as a continuum than as a set of sharply separated classes.

Future work can proceed in several directions. First, methods should be developed to accommodate continuous or graded transitions between groups, because latent subgroups in real disease data may not have discrete boundaries. In particular, extensions based on mixed-membership or latent-variable formulations could allow subjects to be represented as combinations of latent patterns rather than being assigned to a single group, which would provide a more natural description of the AD--MCI--CN disease spectrum and reduce the sensitivity of the results to arbitrary group boundaries. Second, integrating multimodal information, such as genetic data, biomarkers, and clinical covariates, could improve both the identifiability and the interpretability of the estimated groups. When imaging signals alone are insufficient to separate latent subgroups, these additional sources may supply complementary biological and clinical evidence, and they can also be used to characterize the recovered groups in terms of their underlying mechanisms. Practical extensions will need to address the different domains, scales, missing patterns, and acquisition structures across modalities. Third, robust estimation and clustering procedures are needed to handle outliers and heavy-tailed noise, which are common in high-dimensional biomedical data. For example, head motion artifacts (\cite{power2012spurious} and \cite{power2014methods}) and extreme signals induced by various pathological states (\cite{jollans2019quantifying} and \cite{ma2021loss}) can severely distort subject-specific function estimates and clustering. To this end, strategies such as robust loss functions, influence diagnostics, and outlier screening can be employed to mitigate the effects of outliers and heavy-tailed noise on subject-specific function estimation and clustering, thereby improving the reliability of group identification in noisy 3D point-cloud data.

\bibliographystyle{apalike}
\bibliography{bibliography}

\end{document}